\documentclass[twocolumn,showpacs,preprintnumbers,amsmath,amssymb,floatfix]{revtex4-1}
\usepackage{amsmath}
\usepackage{graphicx}
\usepackage{amssymb}
\usepackage{bm}
\usepackage{color}

 \def\be{\begin{equation}}
\def\ee{\end{equation}}
\def\bea{\begin{eqnarray}}
\def\eea{\end{eqnarray}}

\begin{document}

\title{Isoscalar and Isovector spin response in $sd-$ shell nuclei }

\author{H. Sagawa$^{1,2)}$, T. Suzuki$^{3,4)}$, 
}
\affiliation{
$^{1)}$RIKEN, Nishina Center, Wako, 351-0198, Japan\\
$^{2)}$Center for Mathematics and Physics, University of Aizu, Aizu-Wakamatsu, Fukushima 965-8560, Japan\\
$^{3)}$Department of Physics, College of Humanities and Science, 
Nihon Univerity, Sakurajosui 3, Setagaya-ku, Tokyo 156-8550, Japan\\
$^{4)}$National Astronomical Observatory of Japan, Mitaka, Tokyo 181-8588, 
Japan\\
}


\begin{abstract}
The spin magnetic dipole transitions and the neutron-proton spin-spin correlations in $sd-$shell even-even nuclei with $N=Z$ are investigated using shell model wave functions 
taking into accout enhanced  isoscalar (IS) spin-triplet pairing  as well as the effective spin operators.  It was shown that the IS pairing and the effective spin operators gives a large quenching effect on the IV spin transitions to be consistent with  observed data by $(p,p')$ experiments.  
On the other hand, the observed IS spin strength  show much  smaller quenching effect than
expected by the calculated results.  
The IS pairing gives
 a substantial quenching effect on the spin magnetic dipole transitions, especially on the isovector (IV) ones. 
Consequently, an enhanced isoscalar spin-triplet pairing interaction enlarges the proton-neutron spin-spin correlation deduced from the difference between the isoscalar (IS) and the IV sum rule strengths.
The beta-decay rates and the IS magnetic moments of $sd-$shell are also examined in terms of the IS pairing as well as the effective spin operators.  
\end{abstract}

\pacs{21.60.Jz, 21.65.Ef, 24.30.Cz, 24.30.Gd}

\maketitle
\section{Introduction}
The spin-isospin response is a fundamental process in nuclear physics and astrophysics. 
The Gamow-Teller (GT) transition, which is a well-known "allowed"  charge exchange transition, involves the transfer of one unit of the total angular momentum induced by ${\vec \sigma} t_{\pm}$~\cite{Bohr1}. 
In a no-charge-exchange channel, magnetic dipole (M1) transitions are extensively observed in a broad region of the mass table.
Both the spin and the angular momentum operators induce M1 transitions~\cite{Bohr1}, and depending on whether the isospin operator is included also induce the isovector (IV) and the isoscalar (IS) modes.

Compared to the relevant theoretical predictions by shell model and random phase approximations (RPA)~\cite{Gaarde,Wakasa97,Yako05,Yako09,Sasano11,Wakasa12}, the experimental rates of these spin-isospin responses are quenched. 
A similar quenching effect also occurs in the observed magnetic moments of almost all nuclei compared to the single-particle unit (i.e., the Schmidt value)~\cite{Bohr1, Arima,Towner}.
The quenching effect of spin-isospin excitations influences many astrophysical processes such as the mean free path of neutrinos in dense neutron matter, the dynamics and nucleosynthesis in core-collapse supernovae explosions \cite{Lan08}, and the cooling of prototype-neutron stars \cite{Reddy99}.    
Furthermore, the exhaustion of the GT sum rule is directly related to the spin susceptibility of asymmetric nuclear matter \cite{Fan2001} and the spin-response to strong magnetic fields in magnetars \cite{Radhi15}.  

Although the quenching phenomena of magnetic moments and spin responses have been extensively studied, previous research has focused mainly on the mixings of higher particle-hole (p-h) configurations \cite{Arima,Towner,BH} and the coupling to the $\Delta$ resonances \cite{BM-q,MRho}.
In particular, the measured strength of the GT transitions up to the GT giant resonance is strongly quenched compared to the non-energy weighted sum rule, $3(N-Z)$~\cite{Gaarde}.
This observation has raised a serious question about standard nuclear models because the sum rule is independent of the details of the nuclear model, implying a strong coupling to $\Delta$.
After a long debate~\cite{Arima97}, experimental investigations by charge-exchange $(p,n)$ and $(n,p)$ reactions on $^{90}$Zr using multipole decomposition (MD) techniques have revealed about 90\% of the GT sum rule strength in the energy region below E$_{\rm x}$=50~MeV~\cite{Yako05,Ichimura06}, demonstrating the significance of the $2p-2h$ configuration mixings due to the central and tensor forces~\cite{BH}, although the coupling to $\Delta$ is not completely excluded.

IV spin M1 transitions induced by ${\vec \sigma} t_z$ can be regarded as analogous to GT transitions between the same combination of the isospin multiplets.
Therefore, they should show the same quenching effect as GT transitions.
On the other hand, the IS spin M1 transitions are free from the coupling to $\Delta$ and their strength quenching should be due to higher particle-hole configurations.
Various theoretical studies have pointed out that the quenching of IS spin operators is similar to that of IV ones \cite{Richter08}.  
However, recent high-resolution proton inelastic scattering measurements at $E_p$=295~MeV have revealed that the IS quenching is substantially smaller than the IV quenching for several $N$=$Z$ $sd-$shell nuclei~\cite{Matsu2015}.  These empirical findings give rise to a positive value for the proton-neutron spin-spin correlations in the ground state of N=Z nuclei.

Recently, it has been reported that the isoscalar (IS) spin-triplet pairing correlations play an important role in enhancing the GT strength near the ground states of daughter nuclei with mass $N\sim Z$~\cite{Bai3,Fujita14,Tanimura14,Sagawa-Colo}.  
At the same time, the total sum rule of the GT strength is quenched by ground state correlations due to the IS pairing \cite{Zamick02}.

In this paper, we study  the 
IS and  IV spin M1 responses 
 based on modern shell model effective interactions for the same set of $N$=$Z$ nuclei as those in Ref.~\cite{Matsu2015}.   We introduce effective spin and spin-isospin operators 
to make quantitative study of the accumulalted strength in the responses.  
We consider that simultaneous calculations of these responses within the same nuclear model may be advantageous to distinguish the effect of the higher order configurations from  the $\Delta-$hole  coupling due to the fact that the IS spin M1 transition is independent of the $\Delta-$hole coupling strength. We discuss also the effect of IS spin-triplet pairing interaction on the spin responses and the proton-neutron spin-spin correlations in the 
ground states of N=Z nuclei.

The spin M1 operators are introduced in Section II and their sum rules are also defined.
 Section III is devoted to the shell model calculations of N=Z even-even nuclei in comparisons with available experimental data by $(p,p')$ reactions.  The accumulated sum rule values of IS and 
IV spin transitions are extracted in Section 4.  The proton-neutron spin-spin correlations 
are also discussed in terms of the IS spin-triplet pairing correlations.  
The beta-decay rates and IS magnetic moments in $sd-$shell are studied in the same context of the shell model calculations in Section V.  The summary is given in Section VI. 
\section{Spin M1 operators and sum rules}
We consider the IS and IV spin M1 operators, which are given as
\begin{eqnarray}
\hat{O}_{IS}&=&\sum_i{\vec \sigma}(i)  \label{IS-spin},   \\
\hat{O}_{IV}&=&\sum_i{\vec \sigma}(i) \tau_z(i)  \label{IV-spin}, 
\end{eqnarray}
as well as the GT charge exchange excitation operators, which is expressed as
\begin{equation}
\hat{O}_{GT}=\sum_i{\vec \sigma}(i) t_{\pm}(i).
\label{ope-IV-GT}
\end{equation}
The sum rule values for the M1 spin transitions are defined by
\begin{eqnarray}
S(\vec \sigma)=\sum_f\frac{1}{2J_i+1}|\langle J_f||\hat{O}_{IS}||J_i\rangle|^2 ,\\
S(\vec \sigma \tau_z)=\sum_f\frac{1}{2J_i+1}|\langle J_f||\hat{O}_{IV}||J_i\rangle|^2 .
\end{eqnarray}
For the GT transition, the sum rule value is defined by 
\begin{eqnarray}
S(\vec \sigma t_{\pm})=\sum_f\frac{1}{2J_i+1}|\langle J_f||\hat{O}_{GT}||J_i\rangle|^2 ,
\label{GT-sumvalue}
\end{eqnarray}
and satisfies the model independent sum rule,
\begin{equation}
S(\vec \sigma t_-)-S(\vec \sigma t_+)=3(N-Z).
\label{GT-sum}
\end{equation}

According to Ref.~\cite{Matsu2015}, the proton-neutron spin-spin correlation is defined as
\begin{eqnarray}
&&\Delta_{spin}=\frac{1}{16}\left(S(\vec \sigma)-S(\vec \sigma \tau_z)\right)  \nonumber  \\
&=&
\sum_f\langle J_i|\sum_i\frac{{\vec \sigma}_n(i)+{\vec \sigma}_p(i)}{4}|J_f\rangle\langle J_f|\sum_i\frac{{\vec \sigma}_n(i)+{\vec \sigma}_p(i)}{4}|J_i\rangle       \nonumber  \\
&-& \sum_f\langle J_i|\sum_i\frac{{\vec \sigma}_n(i)- {\vec \sigma}_p(i)}{4}|J_f \rangle     
 \langle J_f|\sum_i\frac{{\vec \sigma}_n(i)-{\vec \sigma}_p(i)}{4}|J_i\rangle   \nonumber  \\
&=& \langle J_i|{\vec S_p}\cdot {\vec S_n} |J_i \rangle ,
\label{spin-spin}
\end{eqnarray}
where ${\vec S_p}=\sum_{i\in p}{\vec s_p(i)}$ and  ${\vec S_n}=\sum_{i\in n}{\vec s_n(i)}$.
The correlation value is 0.25 and $-0.75$ for a proton-neutron pair with a pure spin triplet and a singlet, respectively.  
The former corresponds to the  ferromagnet limit of the spin alignment, while the latter is the anti-ferromagnetic one.

\section{Shell model calculations with effective operators and IS pairing correlations}
The shell model calculations are performed in full $sd-$ shell model space with the USDB interaction~\cite{Brown06}. 
Among the effective interactions of the USD family, USD~\cite{Wildenthal}, USDA~\cite{Brown06}, and USDB~\cite{Brown06}, the results of spin excitations with J$^{\pi}$=1$^+$ are quite similar to each other both in excitation energies and transition strengths for collective states with large transition strengths. Hereafter, we present results based on USDB interaction.  
To take into account the effects of higher order configuration mixings as well as meson exchange currents (MEC) and $\Delta-$isobar effect, effective operators are commonly adopted in the study of magnetic moments, GT transitions and spin and spin-isospin dependent $\beta-$decays. 

IV magnetic transitions in $sd-$shell nuclei and GT transitions have been studied extensively
in experiments and theories.   On the other hand, experimental evidence of IS magnetic transitions are not	well known so far except recent experimental data by Matsubara et al..
In the literature \cite{Arima,Towner,BW1983}, the effective operators have been introduced to mimic the effects of higher-order configuration mixings, meson-exchange currents, $\Delta-$isobar coupling and the 
relativistic corrections.  
For the spin operators, the effective operators read for the IS operator
\be
\hat{O}_{IS}^{eff}=f_s^{IS}\vec{\sigma}+f_l^{IS}\vec{l}+f_p^{IS}\sqrt{8\pi}[Y_2\times\vec{\sigma}]^{(\lambda=1)}
\label{eq:IS-eff}
\ee
and also for the IV spin operator,
\be
\hat{O}_{IV}^{eff}=f_s^{IV}\vec{\sigma}\tau_z+f_l^{IV}\vec{l}\tau_z+f_p^{IV}\sqrt{8\pi}[Y_2\times\vec{\sigma}]^{(\lambda=1)}\tau_z
\label{eq:IV-eff}
\ee
where $f_i^{IS(IV)} (i=s,l,p)$ are the effective coefficients of $IS (IV)$ spin, orbital an spin-tensor operators.  The summation of index $i$ in Eq. (1) is discardeed in the effective operators.   The effective coefficients for the IS spin operator obtained by Towner are $f_{s}^{IS}$
=0.745,  $f_{l}^{IS}$ =0.0526 and $f_{p}^{IS}$ =-0.0157.   For the IV part, Towner obtained the corrections for 
the spin, orbital and the spin-tensor operators of GT transitions of $1d-$orbit as
\be
\hat{O}_{GT}^{eff}=(1+\delta g_s)\vec{\sigma}t_{\pm}+\delta g_l\vec{l} t_{\pm}+\delta g_p\sqrt{8\pi}[Y_2\times\vec{\sigma}]^{(\lambda=1)}t_{\pm}
\ee
with
\be
\delta g_s=-0.139, \,\,\,\, \delta g_l=0.0103,\,\,\,\, \delta g_p=0.0283
\ee
due to the various higher order effects.  
 In the shell model calculations with USD interaction, the IV spin and charge exchange GT excitations are the same features since no isospin breaking interaction such as Coulomb interaction and charge symmetry breaking forces is not included.  We adopt the GT  effective operators for IV spin transitions.
For the isoscalar part, the quenching factor for spin operator is introduced to check the sensitivity of transition strength on the effective operator.  The effective operators for IS
orbital and spin-tensor are not introduced in the present study.  For the IS case, 
we take 4 different versions of calculations:
\begin{itemize}
\item
USDB: the original interaction with the bare spin operator
\item
USDB1: the IS spin-triplet pairing matrix is enhanced multiplying a factor 1.1 on the relevant matrix elements of USDB interaction.  The bare spin operator is adopted.
\item
USDB2:the IS spin-triplet pairing matrix is enhanced multiplying a factor 1.1 on the relevant matrix elements of USDB interaction.  The IS spin operator is 10\% 
quenched: $f_s^{IS}$=0.9.
\item
USDB3:the IS spin-triplet pairing matrix is enhanced multiplying a factor 1.2 on the relevant matrix elements of USDB interaction.  The IS spin operator is 10\% 
quenched: $f_s^{IS}$=0.9.
\end{itemize}
For the IV case, we take 3 different cases of calculations:
\begin{itemize}
\item
USDB: the original interaction with the bare spin operator
\item
USDBq1:the IS spin-triplet pairing matrix is enhanced multiplying a factor 1.1 on the relevant matrix elements of USDB interaction.  The effective IV spin operator is adopted.
 \item
USDBq2:the IS spin-triplet pairing matrix is enhanced multiplying a factor 1.2 on the relevant matrix elements of USDB interaction.  The effective IV spin operator is adopted.
\end{itemize}
\subsection{$^{12}$C}
 In $^{12}$C, IS and IV 1$^+$ states are observed at Ex=12.71 and 15.11MeV, respectively.
The B(M1) values are extracted from $(e,e')$ scattering experiments to be 
B(M1)=0.0402 and 2.679 in terms of nuclear magneton $(e\hbar/2mc)^2$ \cite{Cosel2000}.  The shell model calculations with CKPOT interaction give B(M1)=0.01434  and 2.314 in the unit of nuclear magneton at Ex=12.45 and 15.09MeV, respectively,  with the bare magnetic transition operators.  
A recent $p-sd$ shell Hamiltonian, SFO \cite{SFO}, gives B(M1) = 0.0131 and 2.515 $\mu_N^2$ for the IS and IV transitions, respectively.  The model space of SFO is $p-sd$ shell and the excitations from $p-$shell to $sd-$shell are included up to 2$\hbar\omega$. 
It is noticed that the experimental value is about 3 times larger than the calculated value for the IS 1$^+$ state at Ex=12.71MeV, while the calculated value for the IV state is close to the experimental value.  The $(p,p')$ data was reported for the two 1$^+$ states to be B($\sigma)=3.174\pm0.842$ at  Ex=12.71MeV and  B($\sigma\tau)=1.909\pm0.094$ at Ex=15.11MeV, respectively.  
The shell model results with SFO are
B($\sigma)$=1.516 and B($\sigma\tau)$= 1.937, respectively.  The proton inelastic scattering data of  the state at Ex=12.71MeV show also a factor 2 larger value than the shell model results.  The isospin mixing between the two 1$^+$ state has been discussed as an origin of the enhancement of IS spin matrix element.  A large isospin mixing was claimed to 
enhance IS magnetic transition observed by the electron scattering.

 The same effect is expected for B($\sigma$). When the 1$^{+}$, T=0, 12.71 MeV and 1$^{+}$, T=1, 15.11 MeV state are mixed by isospin-mixing, 
\begin{eqnarray}
| 1^{+}, 12.71 MeV> = \sqrt{1-a^2} | 1^{+}, T=0 > +a | 1^{+}, T=1>\nonumber\\
| 1^{+}, 15.11 MeV> = \sqrt{1-a^2} | 1^{+}, T=1 > -a | 1^{+}, T=0>,
\end{eqnarray}
we get an enhancement of B($\sigma$) as well as a reduction of B($\sigma\tau$).
B($\sigma$) is enhanced from 1.516 to 1.714 while B($\sigma\tau$) is reduced from 1.937 to 1.750 
and the mixing amplitude $a$ =0.056 \cite{Flanz}. The difference $\Delta_{spin}$ (Eq. (8)) is found to be enhanced by 0.024.  Though the isospin-mixing gives rise to favorable effects, it is still not enough to reproduce the experimental value of B($\sigma$).  

\subsection{$^{20}$Ne}
\begin{figure}[htp]
\includegraphics[scale=0.3,clip,angle=-90
,bb=0 0 595 842]{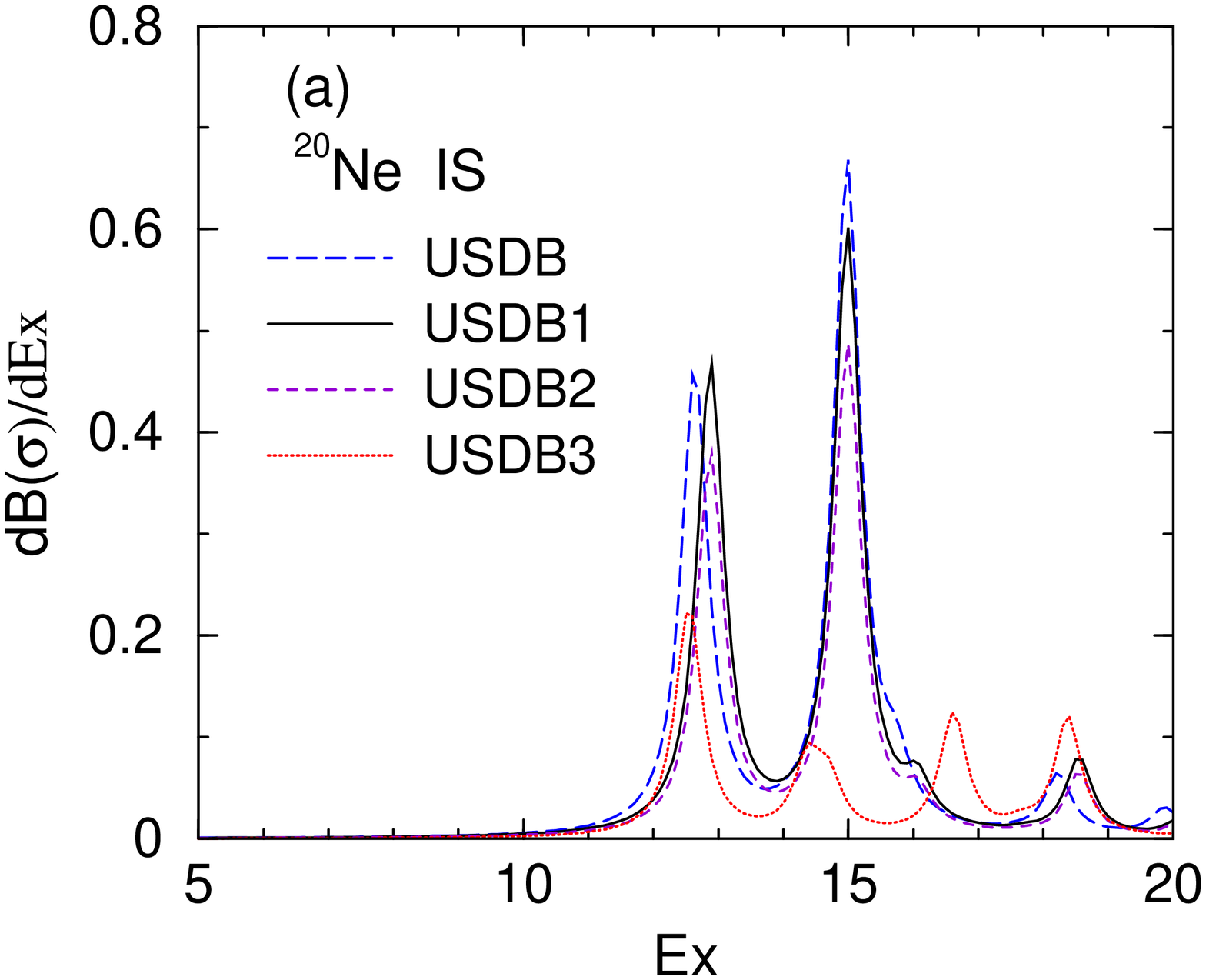}
\includegraphics[scale=0.3,clip,angle=-90
,bb=0 0 595 842]{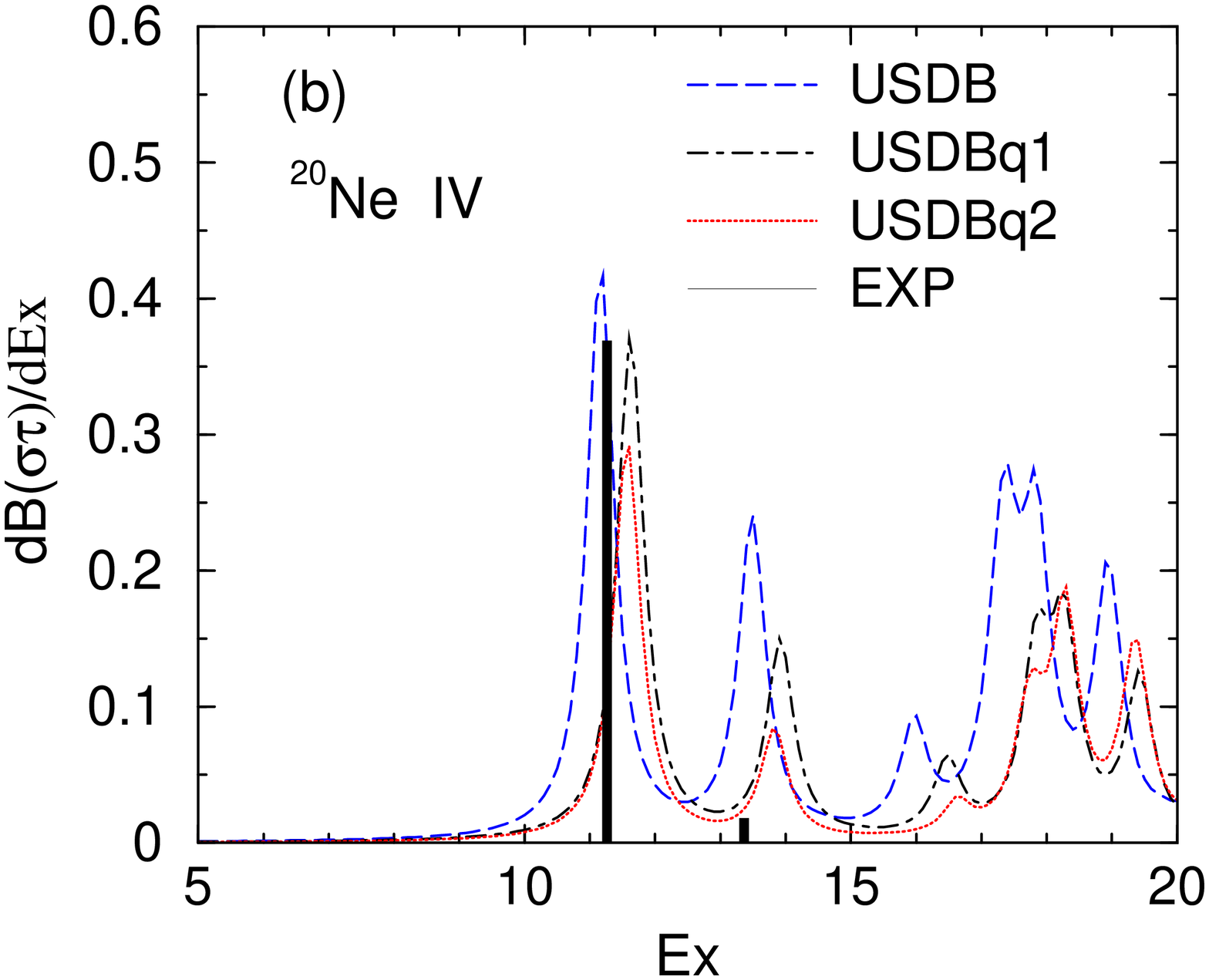}
\caption{(Color online) IS (top) and IV spin-M1 (bottom) transition strengths in $^{20}$Ne.
Shell model calculations are performed in the full $sd-$shell model space with an USDB effective interaction.  
For the IS case, the USDB1 and USDB2 has the 10\% enhanced IS spin-triplet interaction, while USDB3 has 20\% enhanced ones.   The quenching factor for the IS spin operator $f_s^{IS}$=0.9  is introduced for USDB2 and USDB3 calculations. 
For the results of the IV spin-M1 transitions, an effective IV spin operator (11) is adopted in 
USDBq1 and USDBq2 cases.   The  IS spin-triplet interaction is enhanced by multiplying the relevant matrix elements by  factors 1.1  and 1.2 in the cases of USDBq1 and USDBq2, respectively,  together with the effective 
operaror.   
Calculated results are smoothed by taking a Lorentzian weighting factor with the width of 0.5MeV, while the experimental data are shown in the units of B($\sigma$) for the IS excitations and B($\sigma\tau$) for the IV excitations.  
Experimental data are from ref. \cite{Matsu2015}.
\label{fig:ne20-m1}}
\end{figure}

\begin{figure}[htp]
\includegraphics[scale=0.3,clip,angle=-90
,bb=0 0 605 842]{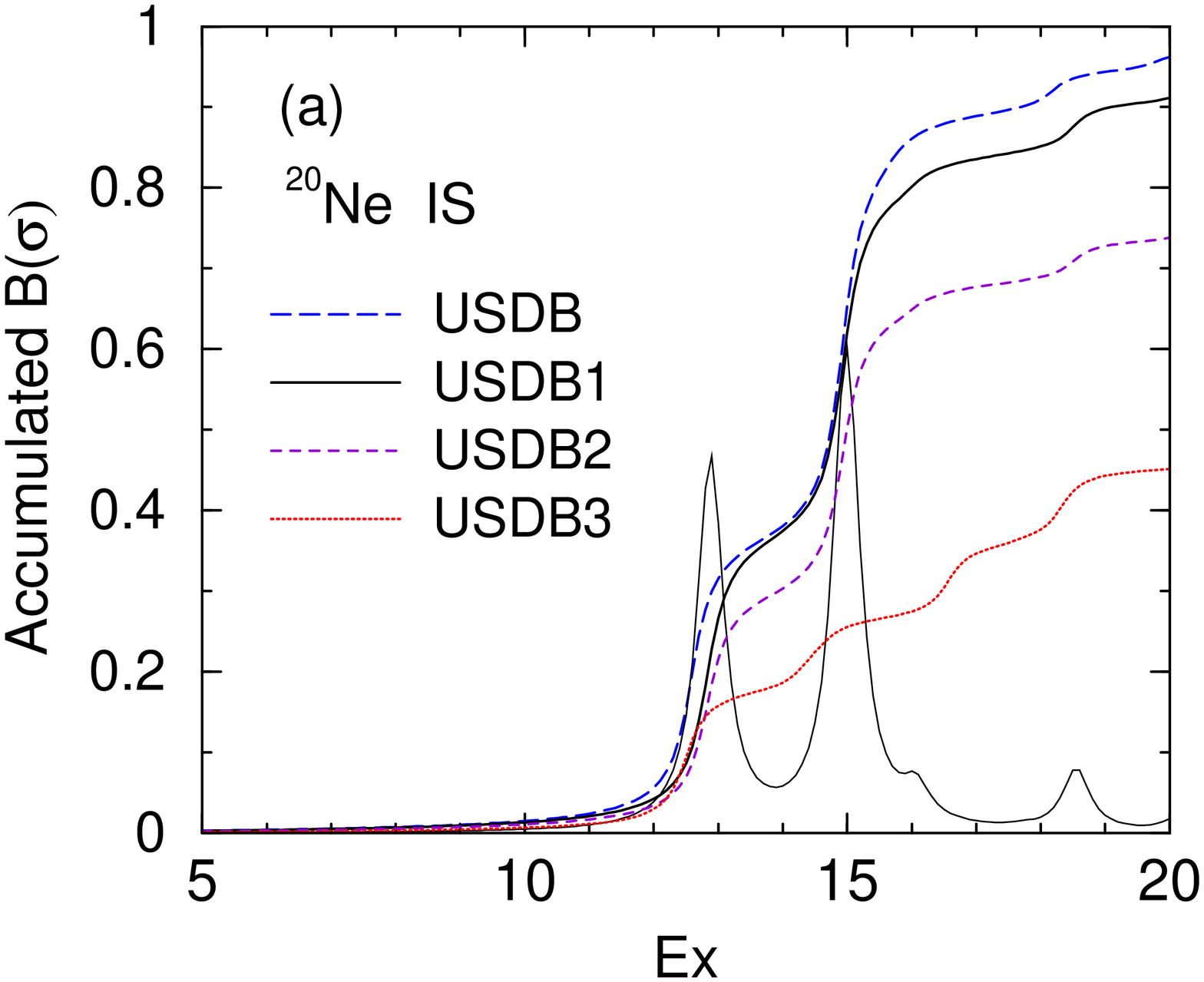}
\includegraphics[scale=0.3,clip,angle=-90
,bb=0 0 595 842]{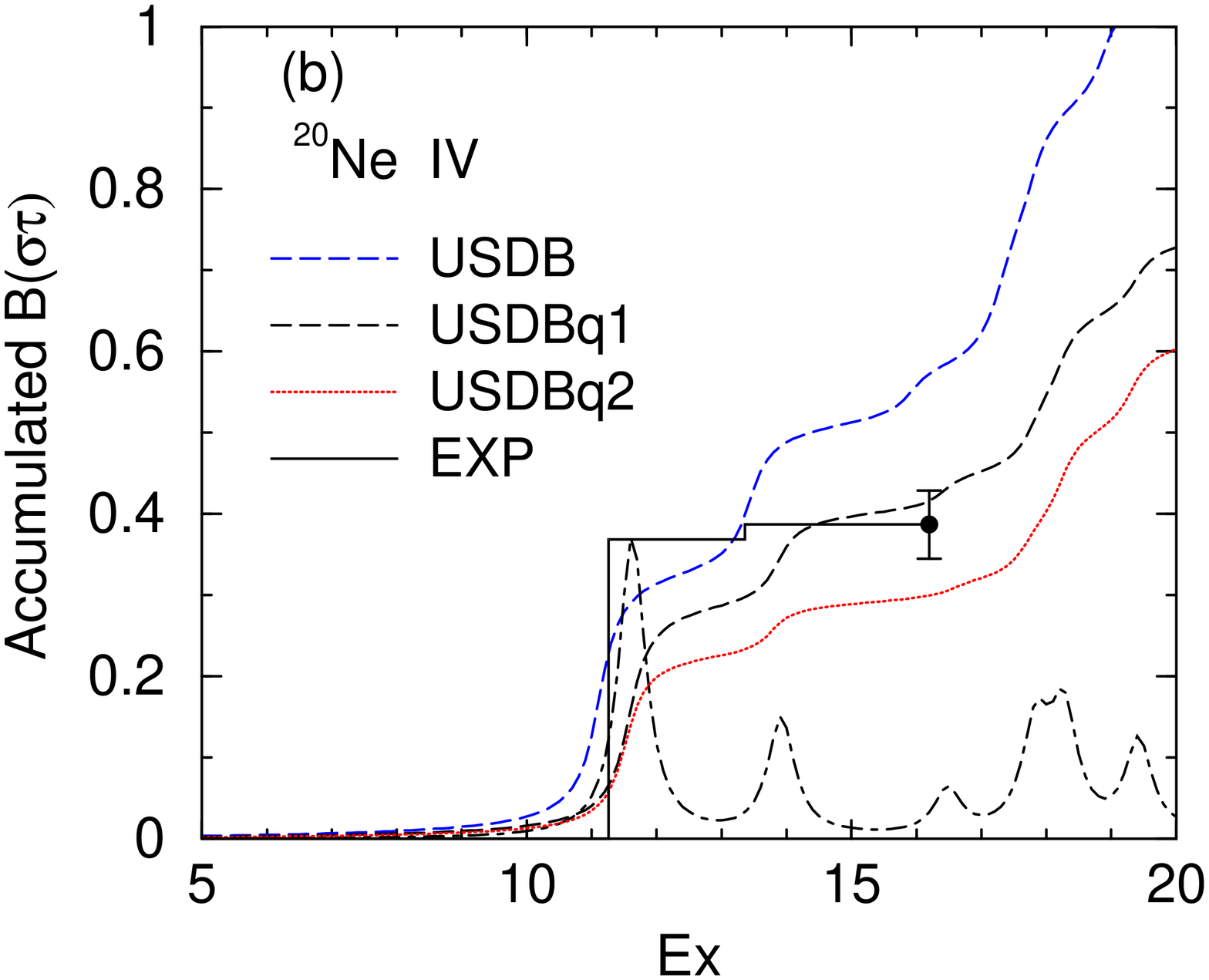}
\caption{(Color online) 
Accumulative sum of the IS spin-M1 strength (top) and the IV spin-M1 strength (bottom) as a function of the excitation energy in $^{20}$Ne. The calculated energy spectra are also shown for USDB1 in the IS channel and for USDBq1 in the IV channel.  
Calculated results are smoothed in the same manner as  Fig.~\ref{fig:ne20-m1}.
Dot with a vertical error bar denotes the experimental accumulated sum of the strengths.
See the text and the caption to Fig, \ref{fig:ne20-m1} for details.  
\label{fig:ne20-m2}}
\end{figure}

Figures \ref{fig:ne20-m1}(a) and \ref{fig:ne20-m2}(a) show the energy spectra of the 
IS spin excitations and their accumulative sums, respectively,  in $^{20}$Ne.  
The IS spin-triplet matrix elements are enhanced by a factor 1.1 for USDB1 and 
USDB2, and by a factor 1.2 for USDB3 case.   Together with the enhancement of IS pairing, the quenched spin operator is introduced in the cases USDB2 and USDB3 with q=0.9.
The calculated results are smoothed by a Lorentzian weighting factor with the width of 0.5 MeV to guide the eye.  
The shell model results with USDB give  1$^+$ states at 
Ex=12.64 and 14.98 MeV with B($\sigma)=0.360$ and 0.519, respectively. 
The USDB3 results with the enhanced pairing and the quenched spin operator  are also shown in the same figure.  The lowest state in the case of USDB3 is found at Ex=12.55MeV with a smaller  strength 
B($\sigma)=0.178$,  which is a half of USDB one.  The higher energy strength is fragmented into three peaks at Ex$\sim$14.5 and 16.6 MeV with the summed strength B($\sigma)$=0.19.  The accumulated values are shown in Fig. \ref{fig:ne20-m2} for four cases USDB, USDB1, USDB2 and USDB2.  
 In $^{20}$Ne, the accumulated sum increases up to Ex$\sim$20MeV.  The enhanced IS pairing in USDB1 gives about 10 \%
quenching compared with USDB results, while more enhanced IS pairing in USDB3 gives further quenching compared with USDB2.  
 In the $(p,p')$ data, the IS strengths are not found so far. 
 
The results of IV spin response are shown in Figures \ref{fig:ne20-m1}(b)  and \ref{fig:ne20-m2}(b).   The original USDB gives IV strength at Ex=11.16 and 13.49MeV with  B($\sigma\tau)$=0.331 and 0.183, respectively, below Ex=15MeV.  The excitation energies of these two peaks are
 shifted to higher energies 11.62 and 13.92MeV in the case of enhanced IS pairing USDBq1.
Comparing with the results of USDBq1, the USDBq2 gives essentially the same excitation energies for 1$^+$ spectra, while the B($\sigma\tau)$ are decreased from 0.292 to 0.230 for  the first peak and from 0.113 to 0.0627 for the second peak.  The calculated results give  large strength also in the energy region above Ex=15MeV.  The accumulated sums are shown in Figure \ref{fig:ne20-m2}(b).
Up to Ex=16MeV, the accumulated sums are 0.577, 0.406 and 0.293 for USDB, USDBq1 and 
USDBq2, respectively.  Due to the strong IS pairing and the effective IS spin operator, the 
accumulated strength decrease by 30\%
 for USDBq1 and 50\% 
for USDBq2.   The $(p,p')$ experiments found two IV strength at Ex =11.26 and  13.36MeV with 
B($\sigma\tau)$=0.369 and 0.018, respectively.  The summed strength 0.387 is comparable with the result of USDBq1.

\subsection{$^{24}$Mg}
\begin{figure}[t]
\includegraphics[scale=0.3,clip,angle=-90
,bb=0 0 595 842]{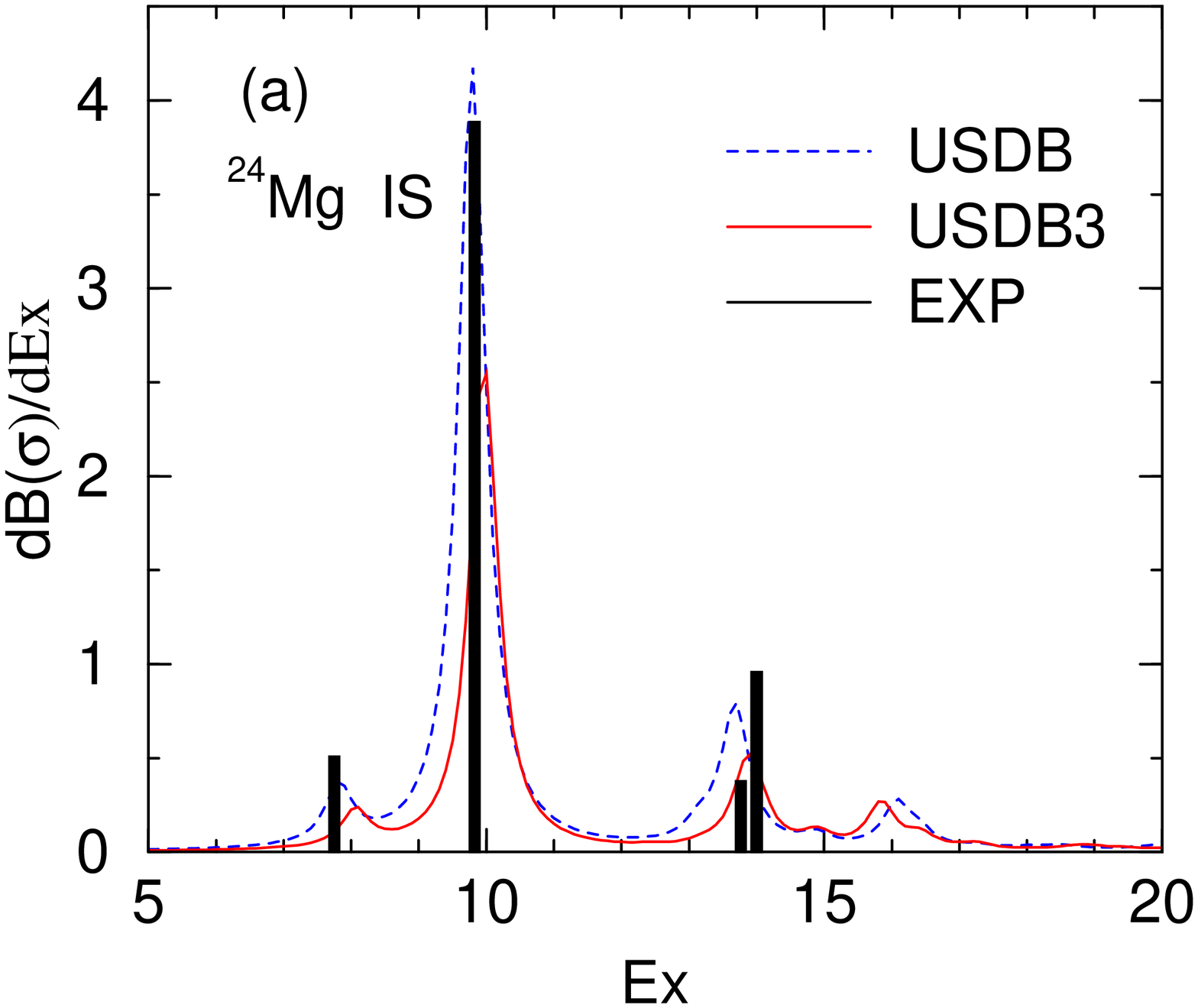}
\includegraphics[scale=0.3,clip,angle=-90
,bb=0 0 595 842]{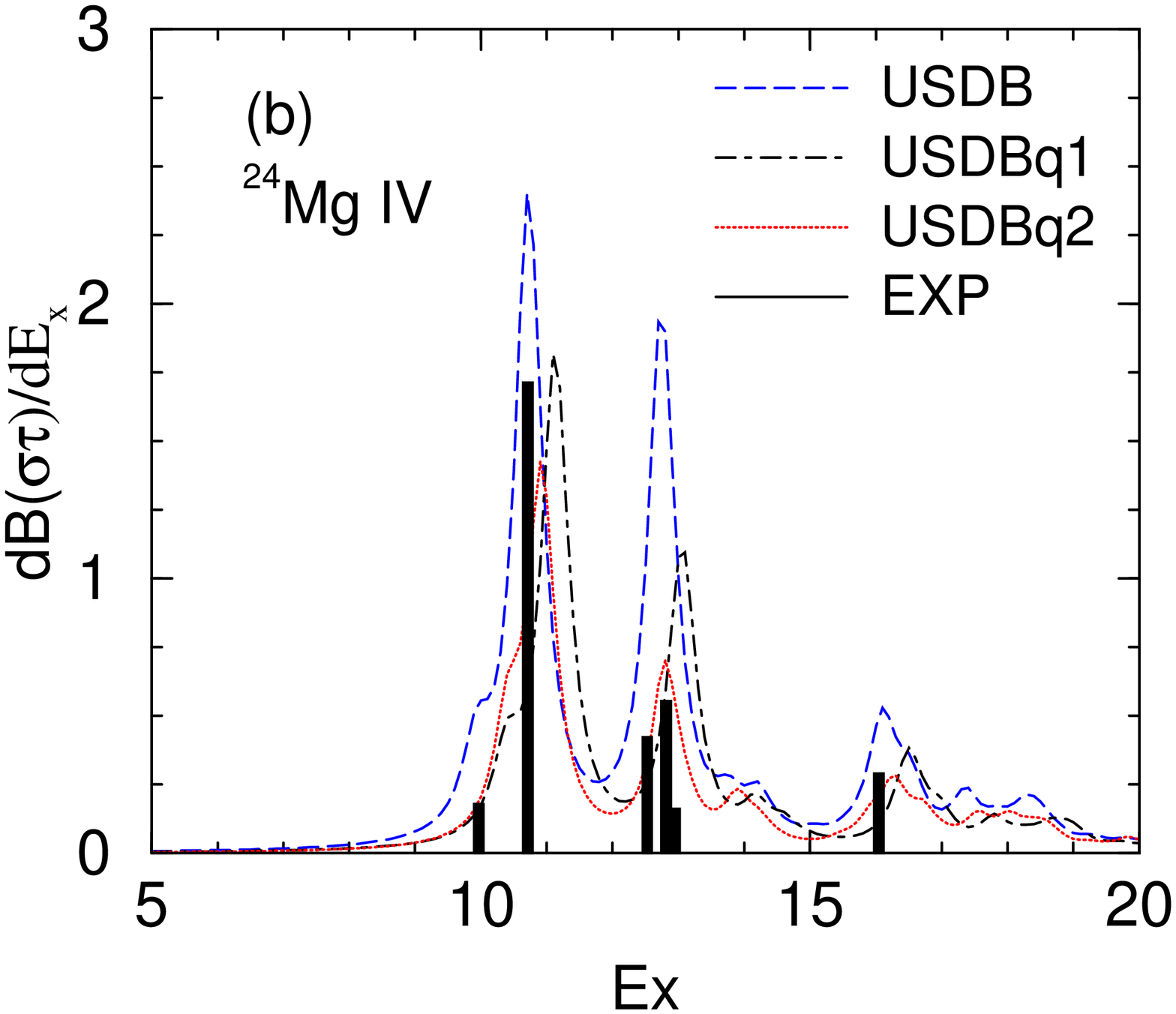}
\caption{(Color online) IS (top) and IV spin-M1 (bottom) transition strengths in $^{24}$Mg.
See the text and the caption to Fig, \ref{fig:ne20-m1} and the text for details.
\label{fig:mg24-m1}}
\end{figure}

\begin{figure}[htp]
\includegraphics[scale=0.3,clip,angle=-90
,bb=0 0 605 842]{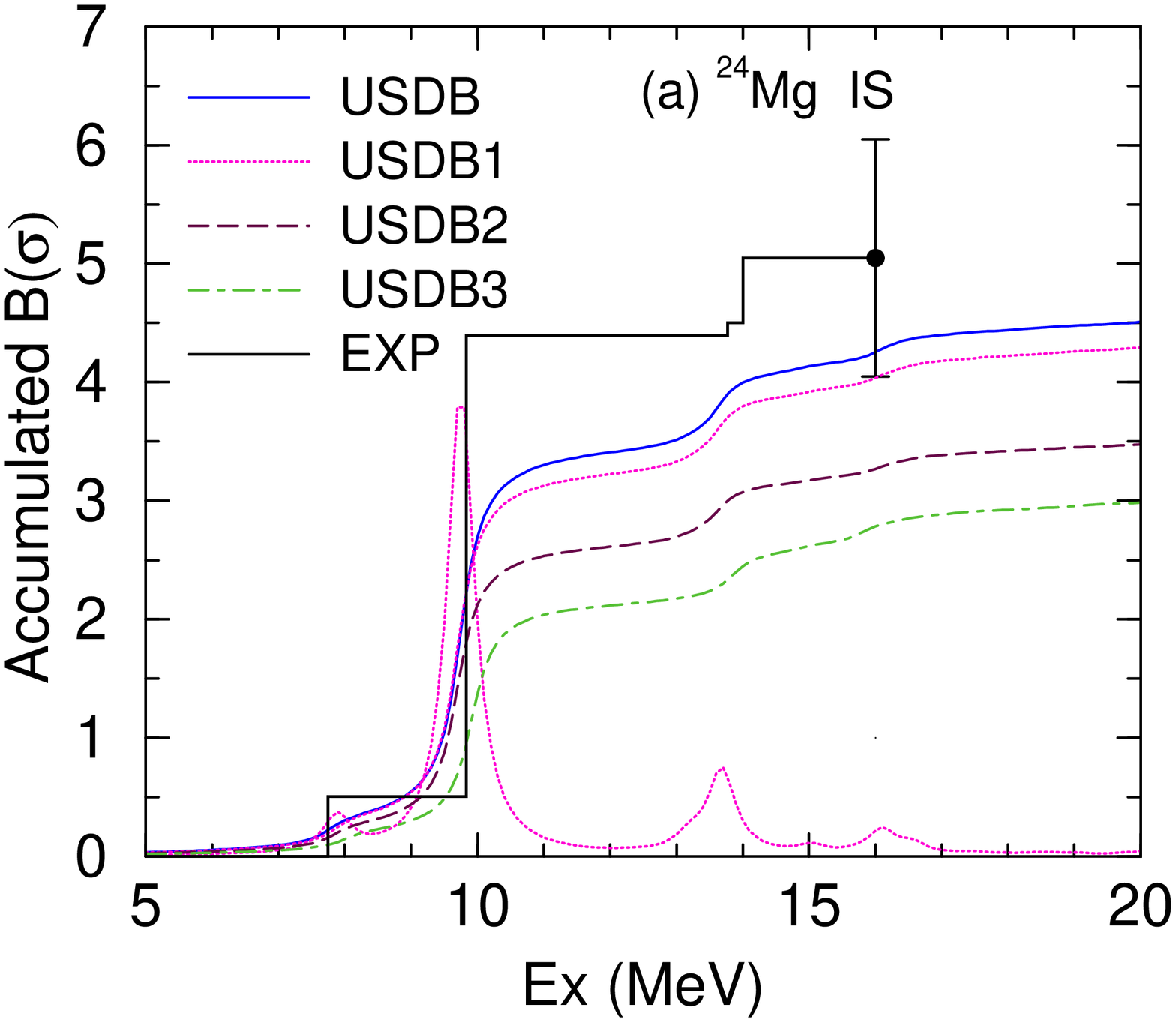}
\includegraphics[scale=0.3,clip,angle=-90
,bb=0 0 595 842]{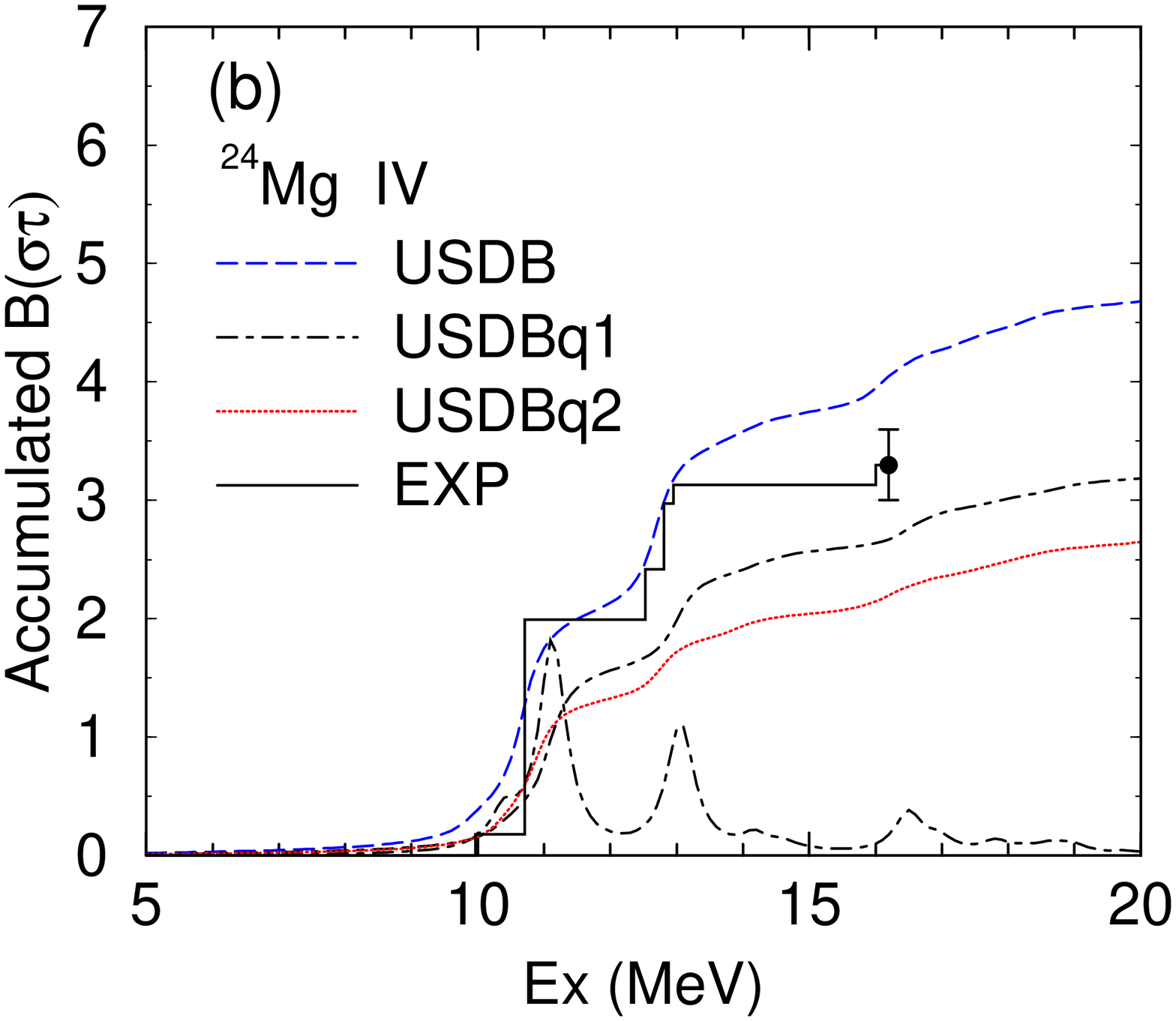}
\caption{(Color online) 
Accumulative sum of the IS spin-M1 strength (top) and the IV spin-M1 strength (bottom) as a function of the excitation energy in $^{24}$Mg.
 See the text and the caption to Fig, \ref{fig:ne20-m2} and the text for details.
\label{fig:mg24-m2}}
\end{figure}

Figures \ref{fig:mg24-m1} and \ref{fig:mg24-m2} show the energy spectra of the 
spin excitations and their accumulative sums, respectively,  in $^{24}$Mg.  
For the IS case, the experimental data give the strong spin M1 strength  at Ex=9.828MeV with B($\sigma)=3.886\pm1.102$.  The shell model results with USDB give IS 1$^+$ state at 
Ex=9.818MeV with B($\sigma)=3.278$.  In the $(p,p')$ data, other IS strengths are also found at 
7.748MeV with B($\sigma)$=0.508 and at Ex$\sim$14MeV with B($\sigma)\sim1.2$.
The calculated results reproduce strong M1 states at  very similar energies Ex=7.82 and 13.7MeV with B($\sigma)=0.24$ and 0.59, respectively.  
The calculations with USDB show also the same amount of  B($\sigma)$ value 
as the experimental data around Ex=14MeV.  
The summed strength up to Ex=16MeV is B$_{exp}(\sigma$:Ex$\leq$16MeV)=5.061$\pm$1.166, while the calculated sum is B$_{cal}(\sigma$:Ex$\leq$16MeV)=4.256.  
The calculated results of USB3 changes only slightly the excitation energies of 1$^+$ state by about 100-200keV, while the summed B($\sigma$) value is decreased by 30\%
.

The experimental analysis show a strong IV spin strength at Ex=10.71MeV with B($\sigma\tau$)=1.714.  The calculation gives at Ex=10.723MeV with B($\sigma\tau$)=1.854.
Experimental data show also substantial strength around Ex=12.8MeV with 
B($\sigma)\sim$1 and at Ex=9.968 and 16.046MeV with B($\sigma\tau)$=0.18 and 0.29, respectively.  The calculated results give large  strengths at Ex=9.939MeV and 10.75MeV with B($\sigma\tau)$=0.238 and 1.521, respectively.  
The experimental summed strength is B$_{exp}(\sigma\tau$:Ex$\leq$16MeV)=3.180$\pm$0.236, 
 while the calculated value is B$_{cal}(\sigma\tau$:Ex$\leq$16MeV)=3.855.  There are about 20\%
quenching in the empirical IV spin sum rule strength below Ex=16MeV compared with USDB results with the bare spin operator.  The USDBq1 and USDBq2 results with the effective spin operator show 
about 30 and 40\% 
quenching of the accumulated strength up to Ex=16MeV, respectively.   

 We study the effects of the isospin-mixing in $^{24}$Mg. 
The 1$^{+}$, T=0 states at E$_{x}$ =7.747 MeV and 9.827 MeV are mixed with the 1$^{+}$, T=1 state at E$_{x}$ = 9.966 MeV \cite{Hoyle1983}. 
Using the mixing amplitudes obtained by $< T=1| V_CD | T=0>/\Delta E$ with $< T=1 | V_CD | T=0>$ = 49 keV \cite{Hoyle1983}, an enhancement of $S(\vec{\sigma})$ from 4.256 to 4.492, a reduction of $S(\vec{\sigma}\tau_z)$ from 3.856 to 3.652 and an enhancement of $\Delta_{spin}$ by 0.026 are obtained for USDB. 
These effects are favorable and consistent with the experimental data though their magnitudes are not significant.   


\subsection{$^{28}$Si}

\begin{figure}[htp]
\includegraphics[scale=0.3,clip,angle=-90
,bb=0 0 595 842]{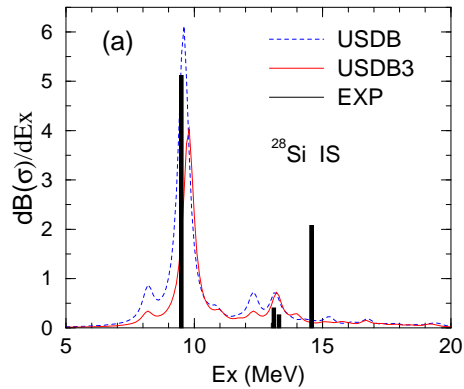}
\includegraphics[scale=0.3,clip,angle=-90
,bb=0 0 595 842]{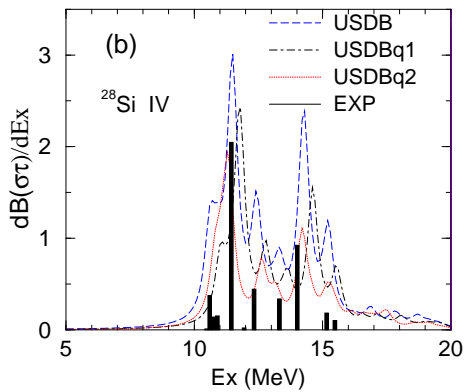}
\caption{(Color online) IS (top) and IV spin-M1 (bottom) transition strengths in $^{28}$Si.
See captions to Fig. \ref{fig:ne20-m1} and the text for details. \label{fig:si28-m1}}
\end{figure}

\begin{figure}[t]
\includegraphics[scale=0.3,clip,angle=-90
,bb=0 0 605 842]{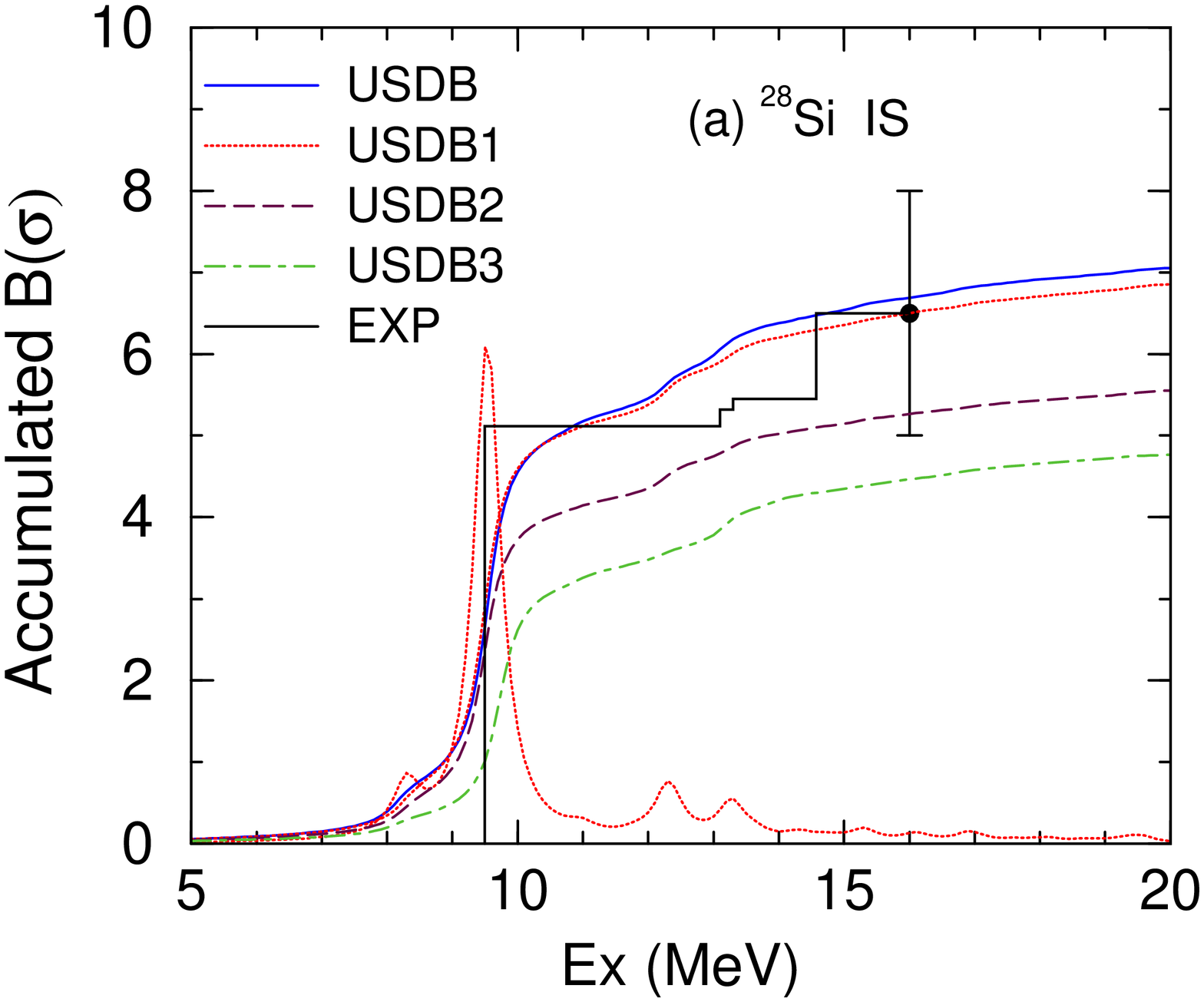}
\includegraphics[scale=0.3,clip,angle=-90
,bb=0 0 595 842]{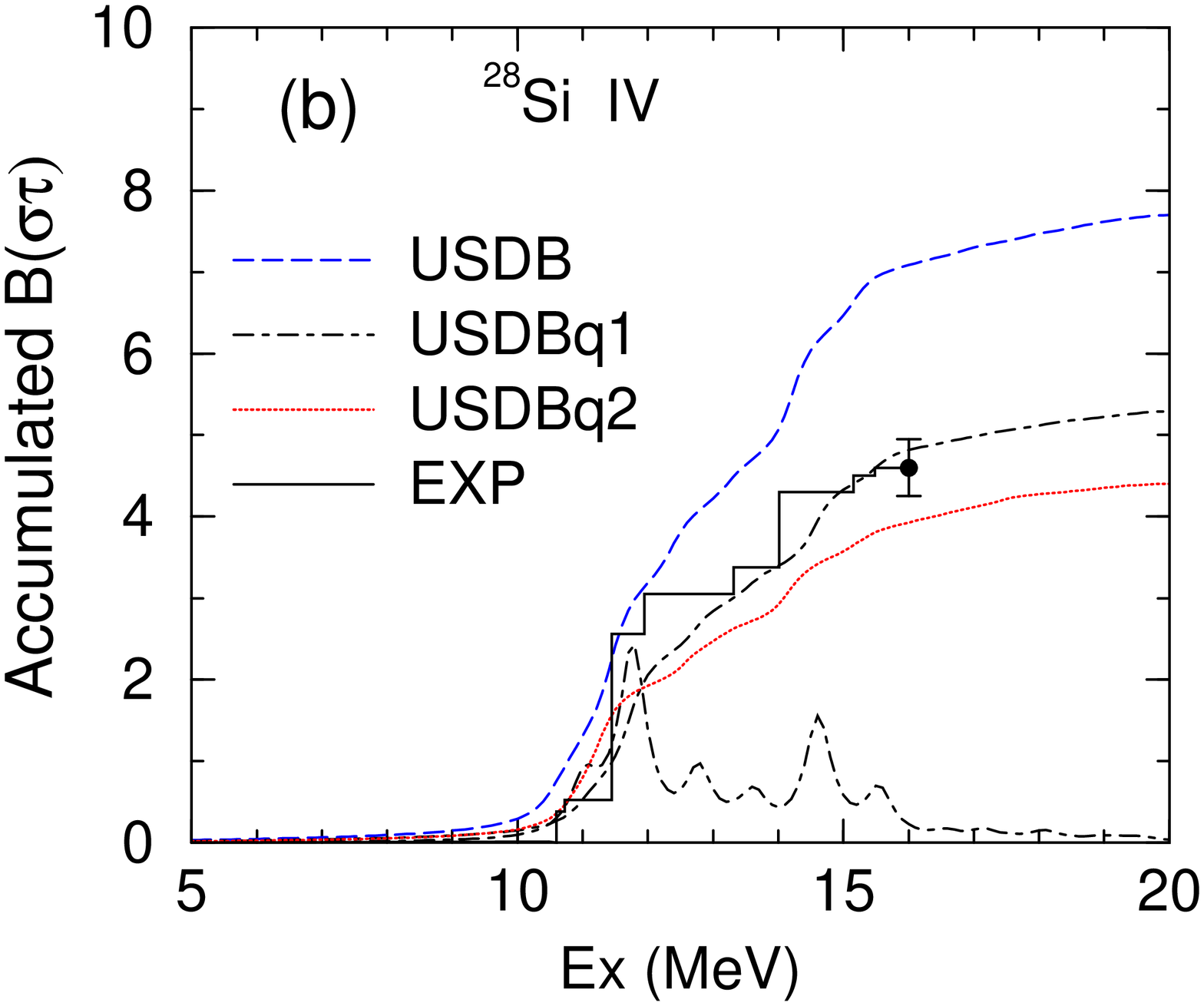}
\caption{(Color online) 
Accumulative sum of the IS spin-M1 strength (top) and the IV spin-M1 strength (bottom) as a function of the excitation energy in $^{28}$Si.
See the captions to Fig.  \ref{fig:ne20-m2}  and the text  for details.
\label{fig:si28-m2}}
\end{figure}

The calculated results of  IS response are shown in  Figs. \ref{fig:si28-m1}(a) and \ref{fig:si28-m2}(a).   The calculated results give a IS 1$^+$ state at Ex=9.6MeV with USDB 
interaction,  which reproduces  well the experimental IS 1$^+$ state with a strong spin transition at E$_{\rm{x}}$=9.58MeV exhausting  about 70\% of the total IS strength.
Another strong state is observed at Ex=14.571MeV with  B($\sigma)$=2.075$\pm$0.621, 
while the calculation shows no sign of strong B($\sigma)$ transition above Ex=14MeV.
Other IS spin transitions are found experimentally at Ex$\sim$13MeV with B($\sigma)\sim$0.7.  The calculations show also the IS spin strength of B($\sigma)\sim$1.0 at Ex=(12.3-13.2)MeV.  The summed empirical IS strength below Ex=16MeV is  
B$_{exp}(\sigma$:Ex$\leq$16MeV)=6.489$\pm$1.604,  while  the calculated results are B$_{cal}(\sigma$:Ex$\leq$16MeV)=6.816,  6.611,  5.355 and 4.562 for USDB, USDB1, USDB2 and
USDB3, respectively.  The  B$_{cal}(\sigma$:Ex$\leq$16MeV) values show  3, 21 and 33\%
quenching for USDB1, USDB2 and
USDB3 interactions, respectively, compared with USDB results.


The IV spin response is shown in Figs.~\ref{fig:si28-m1}(b) and  \ref{fig:si28-m2}(b).
 Gross structure of empirical IV spin response is well reproduced by the calculations based on USDB interaction.
Empirical IV spin strength is rather fragmented,  while 
two IV 1$^+$ states with strong spin strengths of B$(\sigma\tau)$=2.05 and 0.92  are reported at  E$_{\rm{x}}$=11.45 and 14.01MeV, respectively.  Calculated results show also 
largely fragmented IV spin strength and two  strong strength are found at Ex=11.48 and 14.26MeV with 
B$(\sigma\tau)$=2.165 and 1.773, respectively, with USDB interaction.
The empirical IV summed strength is B$_{exp}(\sigma\tau$:Ex$\leq$16MeV)=4.59$\pm$0.222, while the calculated one is 
B$_{cal}(\sigma\tau$:Ex$\leq$16MeV)=7.34, 5.03 and 4.03 for USDB, USDBq1 and USDBq2 cases, respectively.  In the spin IV sum rule value, we see a
large quenching spin factor q$_{s}^{IV}$(eff)$\equiv\sqrt{B_{exp}(\sigma\tau)/ B_{cal}(\sigma\tau:\rm{USDB})}$=0.79,  which is close to the ratio of summed values of USDBq1 to USDB.   
It is pointed out also in ref. \cite{Sagawa2016} that the enhanced IS pairing multiplying a factor 1.2 on the IS pairing matrices reduces the IV spin transition strength, corresponding to the renormalization factor of $f_{s}^{IV}$=0.87  for the accumulated IV spin strength. 
On the other hand, the same enhanced IS pairing gives the IS quenching factor $f_{s}^{IS}$=0.91.
This difference between IS and IV spin response induces a positive value for the proton-neutron spin-spin correlations in the ground state.  This point will be discussed more in Section  IV.  
 


\subsection{$^{32}$S}
\begin{figure}[t]
\includegraphics[scale=0.3,clip,angle=-90
,bb=0 0 595 842]{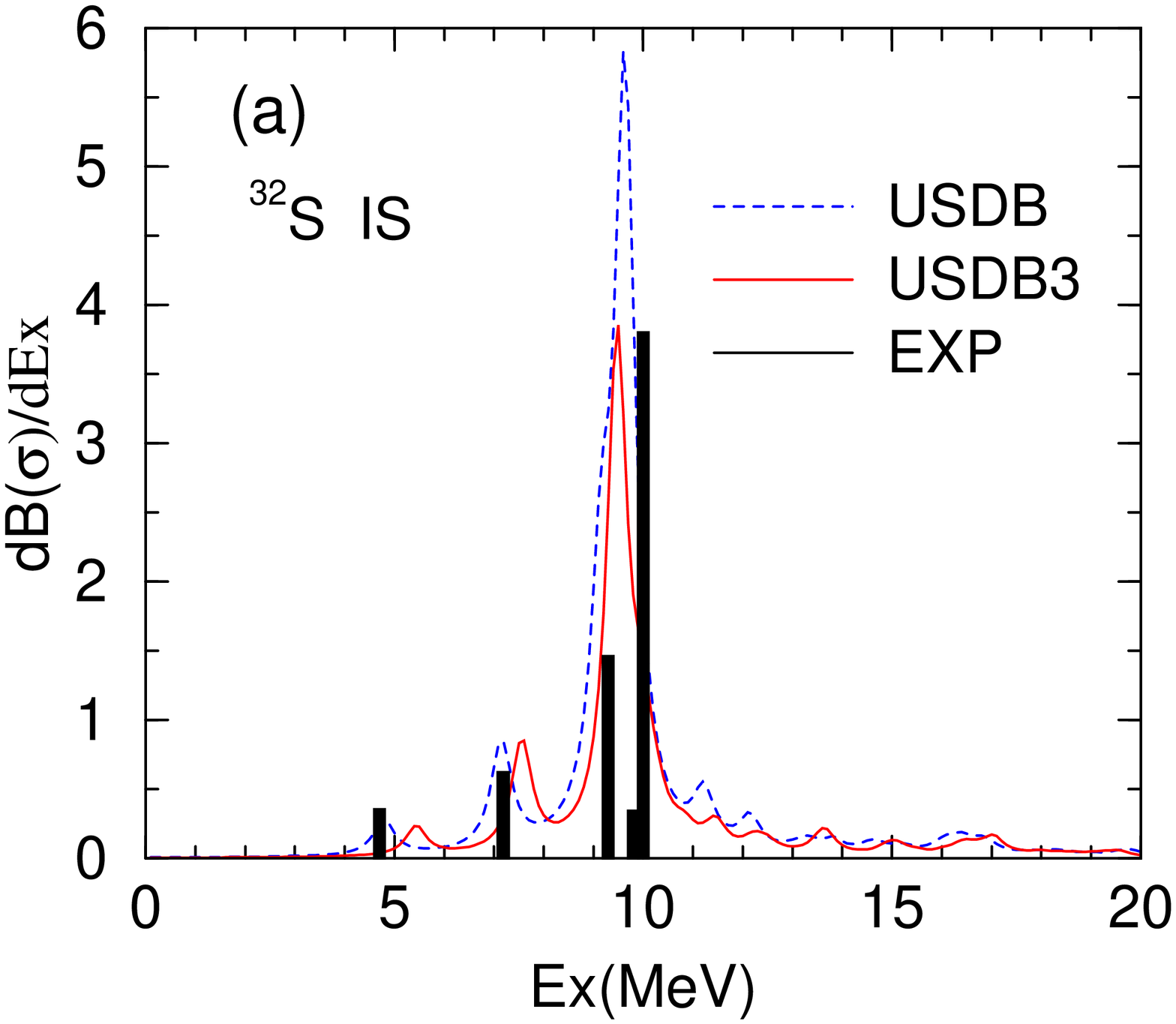}
\includegraphics[scale=0.3,clip,angle=-90
,bb=0 0 595 842]{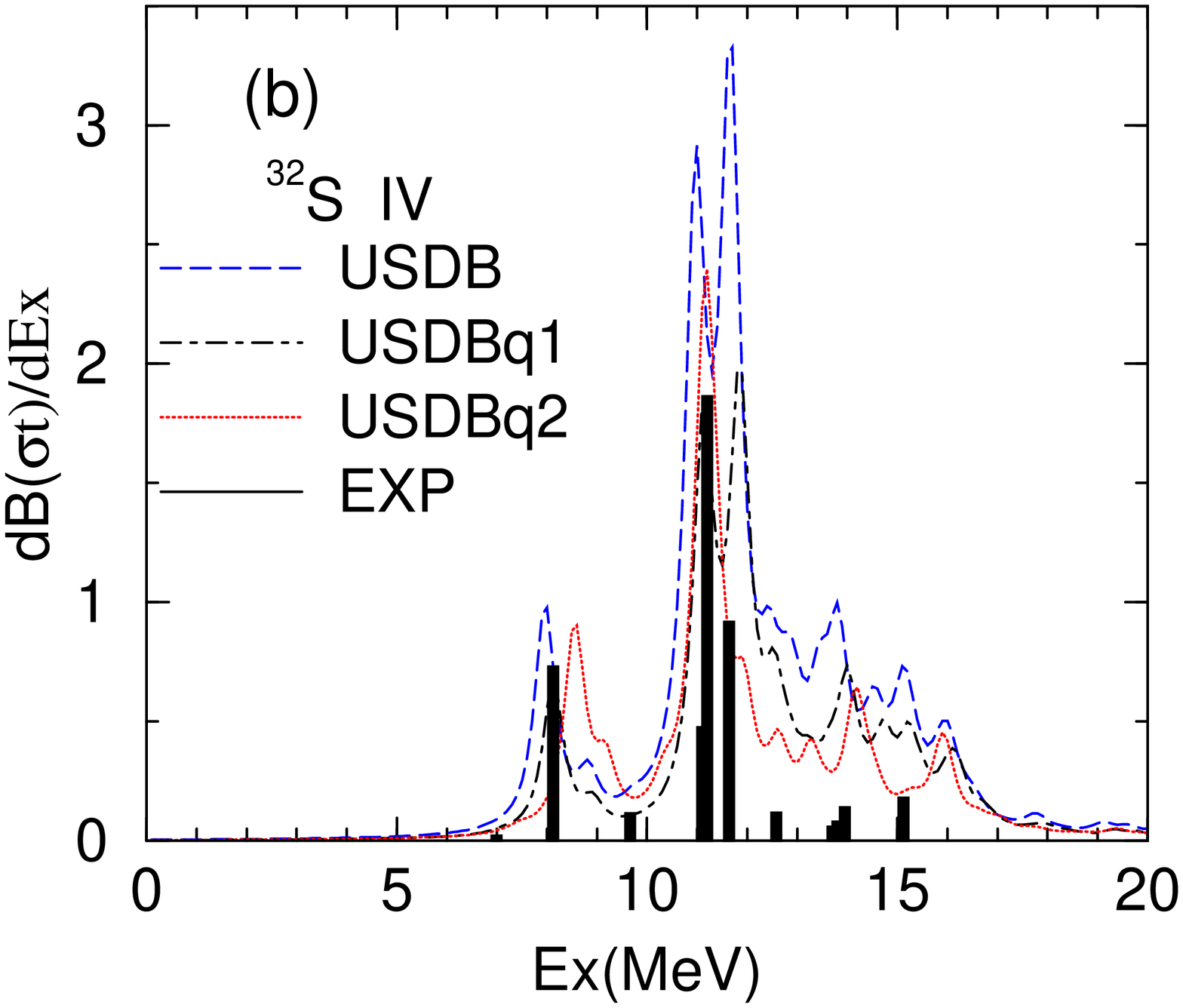}
\caption{(Color online) IS (top) and IV spin-M1 (bottom) transition strengths in $^{32}$S.
See captions to Fig. \ref{fig:ne20-m1} for details. \label{fig:s32-m1}}
\end{figure}

For the IS case shown in Fig. \ref{fig:s32-m1}(a), the experimental data show a strong state at Ex=9.956MeV with B($\sigma)=3.810\pm1.118$.  The corresponding state is found in the calculated results at Ex=9.632MeV with B($\sigma)=$4.312.  Another strong IS transition was found at Ex=9.297MeV with B($\sigma)=1.461\pm0.436$, while the calculations show a state at
Ex=9.154MeV with   B($\sigma)=$1.293MeV.  There are two IS states observed below Ex=7.2MeV. The calculations found also two states at the same energy region with almost the same  B($\sigma$) values as the observed ones.   
The observed IS sum rule strength is 
B$_{exp}(\sigma$:Ex$\leq$16MeV)=6.414$\pm$1.227,  while   theoretically B$_{cal}(\sigma$:Ex$\leq$16MeV)=7.623.  We can see a small quenching effect corresponding to  
$f_{s}^{IS}(eff)$=0.92 for the sum rule strength below Ex=16MeV.

The IV response in $^{32}$S is shown Fig. \ref{fig:s32-m1}(b).  The IV spin strength is concentrated at Ex$\sim$11.3MeV having 80\%
of the total strength below Ex=16MeV.  The calculated results show also very large fraction of the total strength of about 87\%
of the total strength.  Another strong state is found experimentally at Ex=8.125MeV with 
B($\sigma\tau)=0.730\pm0.040$, while the calculations show a state at Ex=7.959MeV with 
B($\sigma\tau$)=0.743.  The agreement between theory and experiment is  quite
satisfactory as far as the gross feature of IV spin response is concerned.  The summed strength of IV transitions is 
B$_{exp}(\sigma\tau$:Ex$\leq$16MeV)=4.120$\pm$0.407, while the calculated results are
B$_{exp}(\sigma\tau$:Ex$\leq$16MeV)=7.993.   We see a large quenching for IV case with
$f_{s}^{IV}(eff)$=0.72.  The results USDBq2 with the enhanced IS pairing and the effective IV spin operator give good account of the accumulated strength.

\begin{figure}[htp]
\includegraphics[scale=0.3,clip,angle=-90
,bb=0 0 605 842]{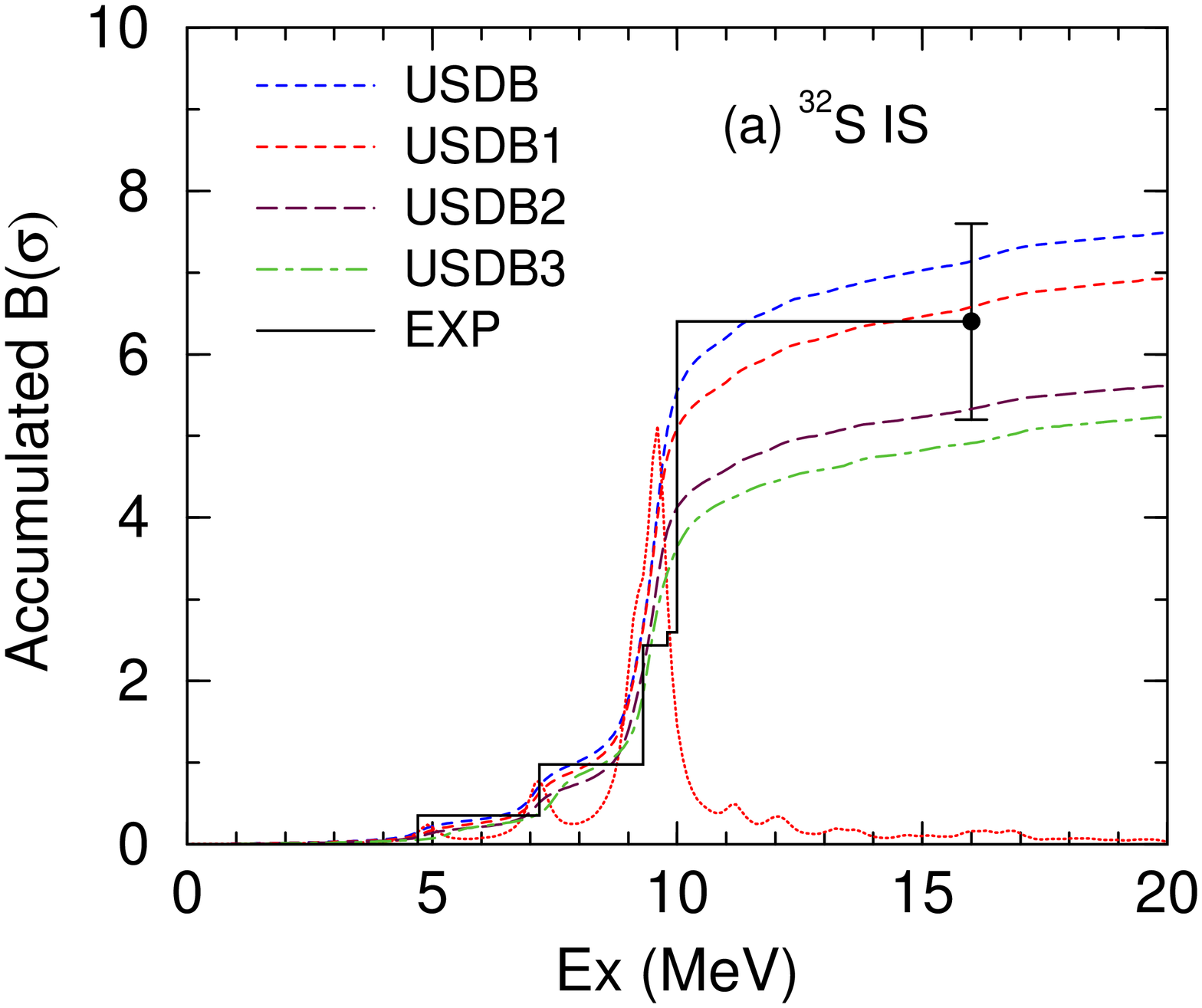}
\includegraphics[scale=0.3,clip,angle=-90
,bb=0 0 595 842]{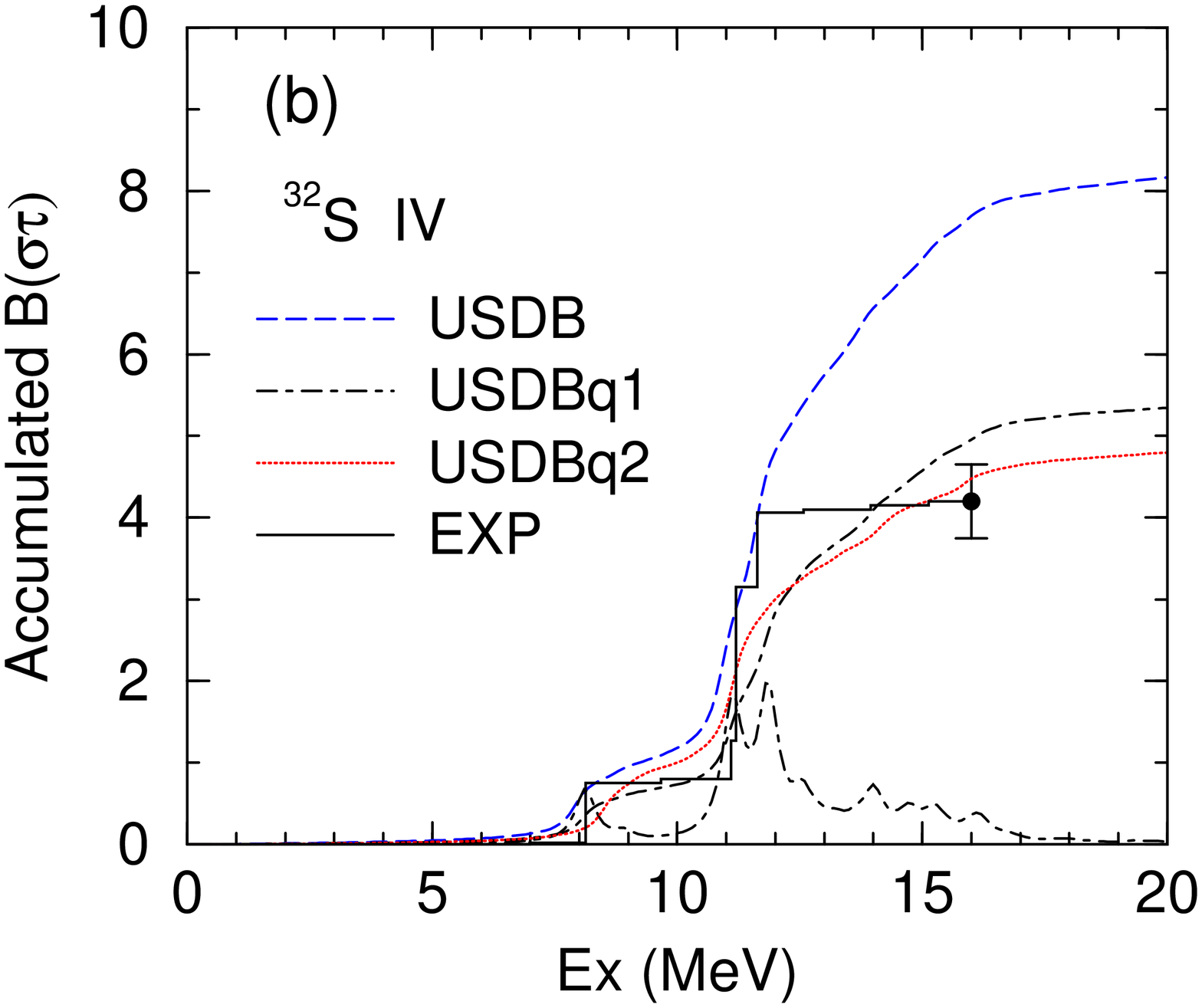}
\caption{(Color online) 
Accumulative sum of the IS spin-M1 strength (top) and the IV spin-M1 strength (bottom) as a function of the excitation energy in $^{28}$Si.
See the captions to Fig.  \ref{fig:ne20-m2} for details.
\label{fig:s32-m2}}
\end{figure}


\subsection{$^{36}$Ar}
\begin{figure}[htp]
\includegraphics[scale=0.3,clip,angle=-90
,bb=0 0 595 842]{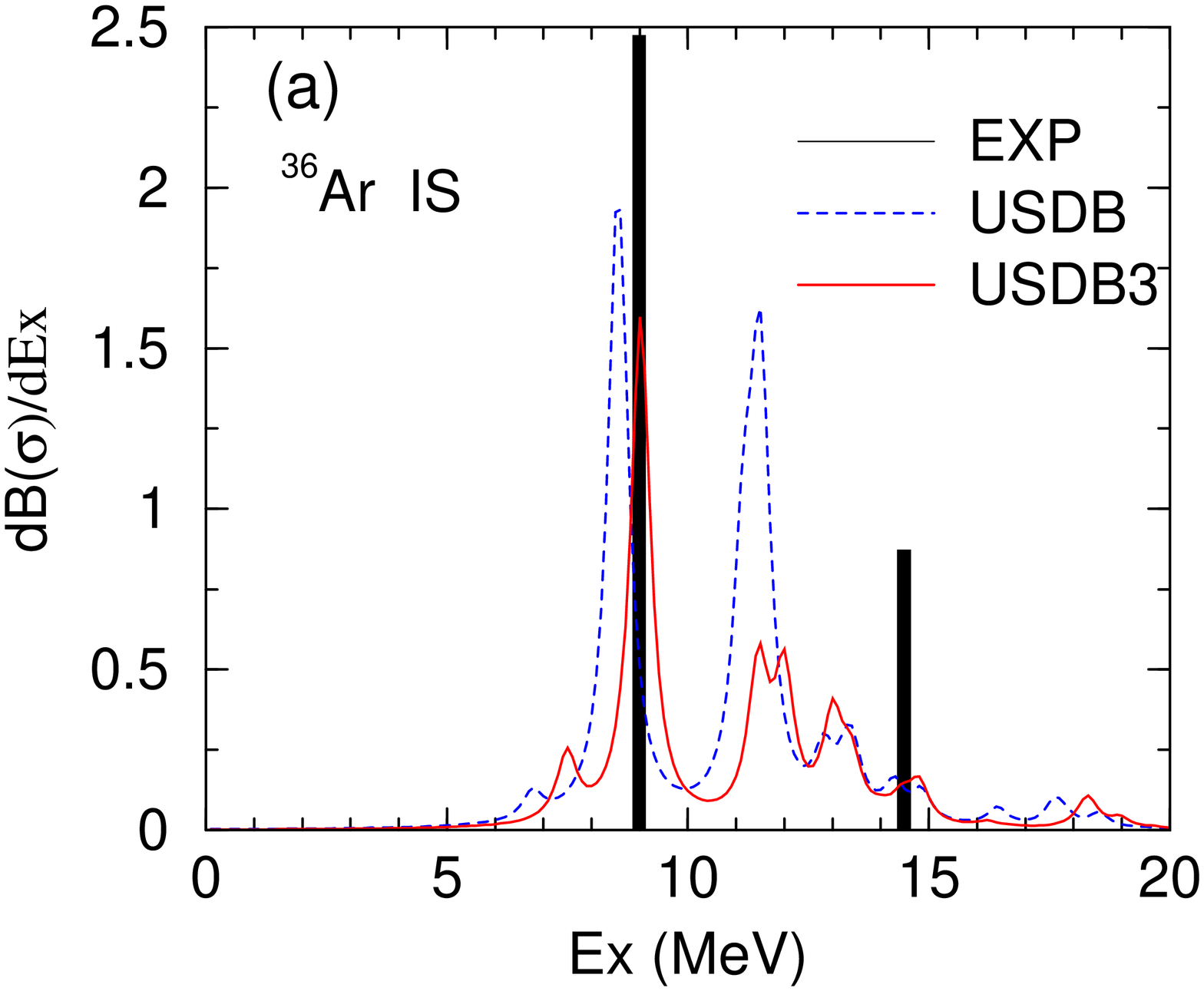}
\includegraphics[scale=0.3,clip,angle=-90
,bb=0 0 595 842]{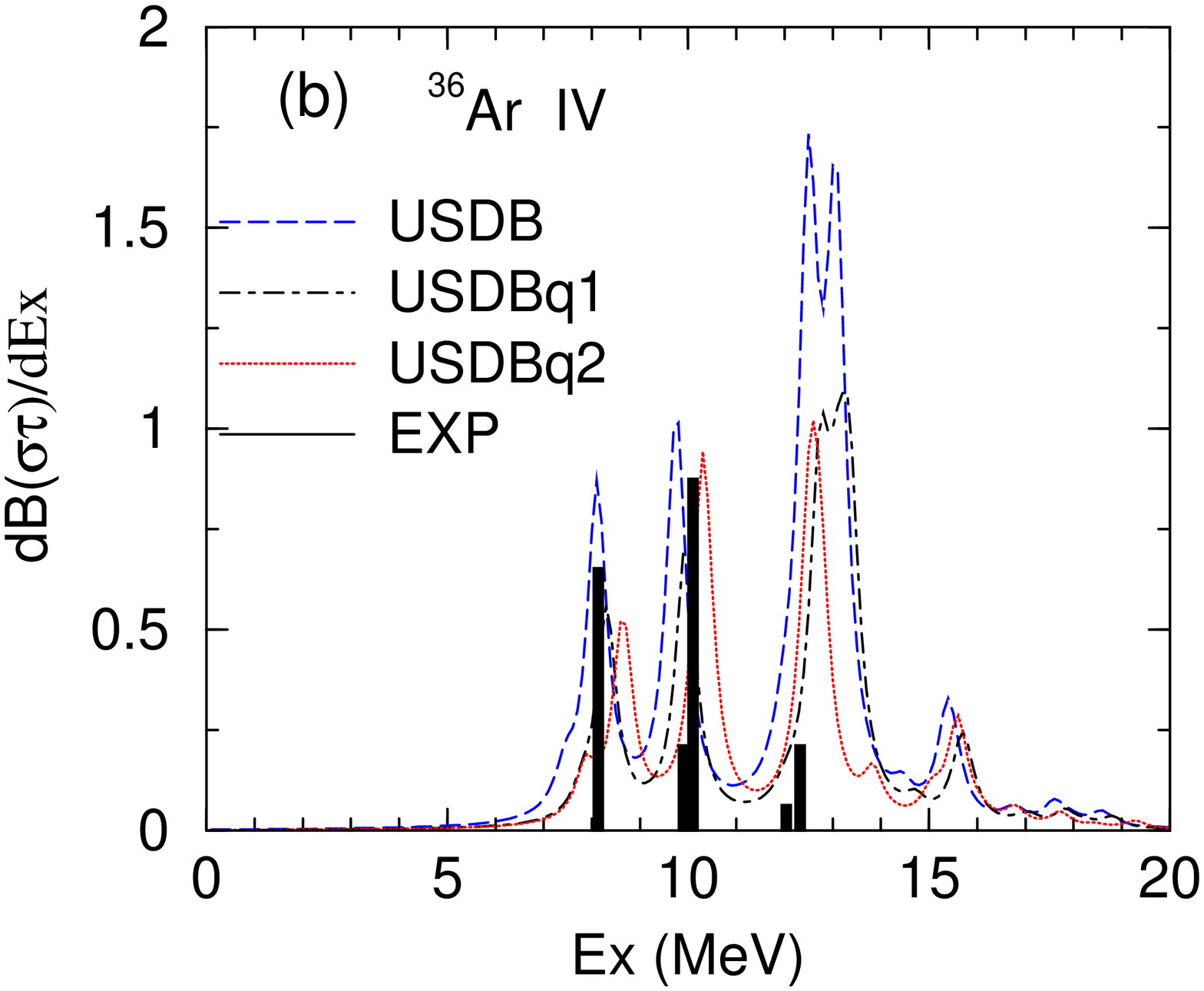}
\caption{(Color online) IS (top) and IV spin-M1 (bottom) transition strengths in $^{36}$Ar.
See captions to Fig. \ref{fig:ne20-m1} for details. \label{fig:ar36-m1}}
\end{figure}

\begin{figure}[htp]
\includegraphics[scale=0.3,clip,angle=-90
,bb=0 0 605 842]{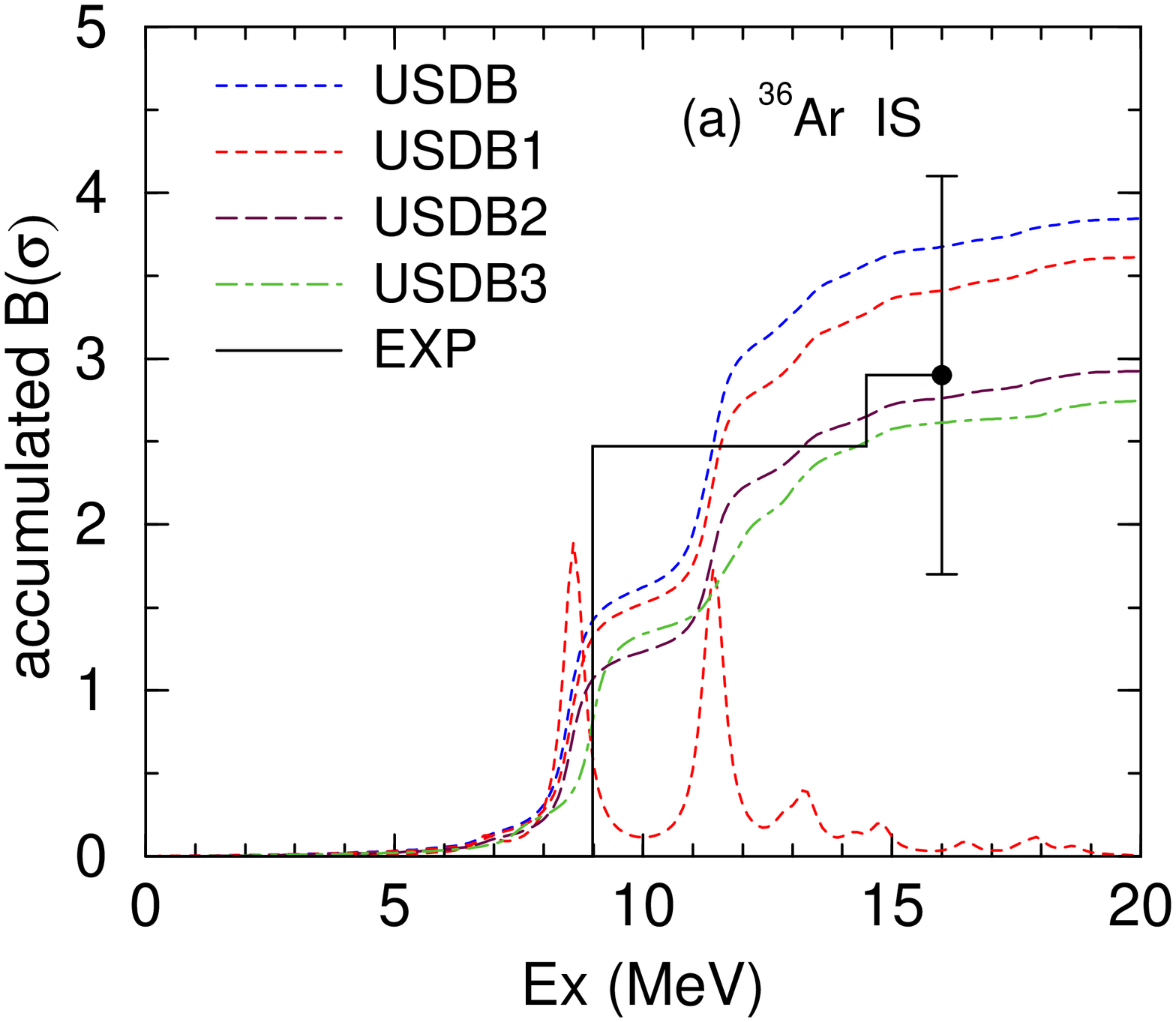}
\includegraphics[scale=0.3,clip,angle=-90
,bb=0 0 595 842]{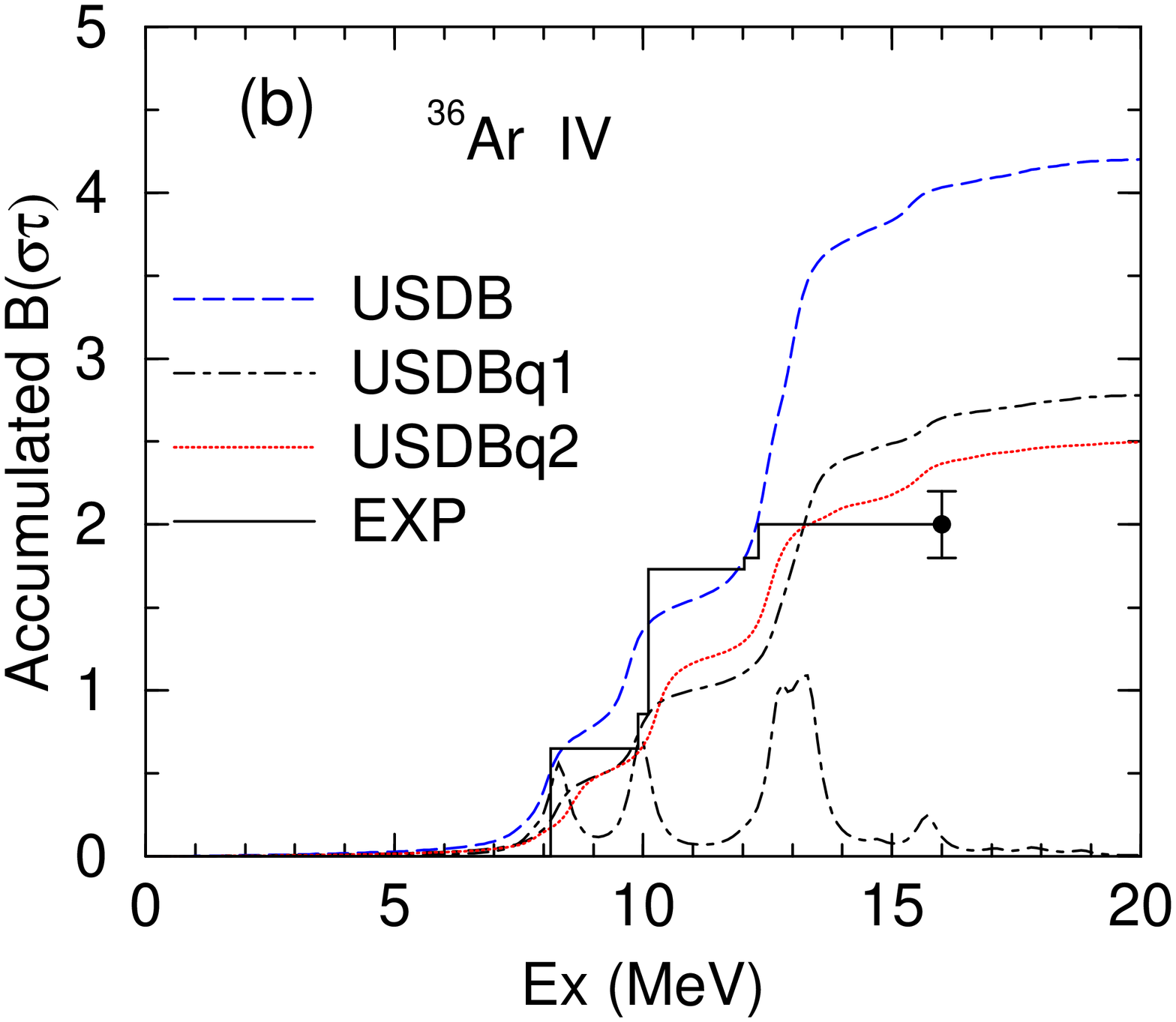}
\caption{(Color online) 
Accumulative sum of the IS spin-M1 strength (top) and the IV spin-M1 strength (bottom) as a function of the excitation energy in $^{28}$Si.
See the captions to Fig.  \ref{fig:ne20-m2} for details.
\label{fig:ar36-m2}}
\end{figure}

The IS spin response in $^{36}$Ar is given in Fig.  \ref{fig:ar36-m1}(a).  The experiments are found two states at Ex=8.985 and 14.482MeV with B($\sigma$)=2.473$\pm$1.045 and
0.872$\pm$0.342, respectively.  The shell model results of USDB show a strong IS strength at Ex=8.551MeV with B($\sigma$)=1.558.  Above Ex=10MeV, the calculated IS strength is rather fragmented with the summed  B($\sigma)\sim$2 in the energy region  Ex=(11-15)MeV.  The experimental summed strength  in Fig.  \ref{fig:ar36-m2}(a) is  
B$_{exp}(\sigma$:Ex$\leq$16MeV)=2.910$\pm$1.091, while the calculated value is
B$_{cal}(\sigma$:Ex$\leq$16MeV)=3.753. We see a small quenching with the factor
 f$_{s}^{IS}$(eff)=0.88.

The IV spin strength is shown in Fig.  \ref{fig:ar36-m1}(b).  Experimental data show two large  IV strength at Ex=8 and 10MeV having 33\%
 and 55\% 
of the observed total strength below Ex=16MeV.  The calculated USDB results show also two strong spin strengths at Ex=8.11MeV and Ex=9.8MeV with B($\sigma\tau$)=0.648 and 0.788, respectively.   The calculations show another strong IV transition strength with B($\sigma\tau)\sim$2.3 at Ex$\sim$13MeV, while experimental data observed a small strength with B($\sigma\tau)\sim$0.3 around Ex=12MeV.  The observed summed strength in Fig.  \ref{fig:ar36-m2}(b)  is B$_{exp}(\sigma\tau$:Ex$\leq$16MeV)=1.986$\pm$0.143, while the calculated one is B$_{cal}(\sigma\tau$:Ex$\leq$16MeV)=4.125.  The quenching factor is rather large with  $f_{s}^{IV}$(eff)=0.69 for the IV case so that the results of USDBq2 in  Fig.   \ref{fig:ar36-m2}(b)  give the closest value to the empirical one.

\begin{figure}[t]
\includegraphics[scale=0.3,clip,angle=-90
,bb=0 0 595 842]{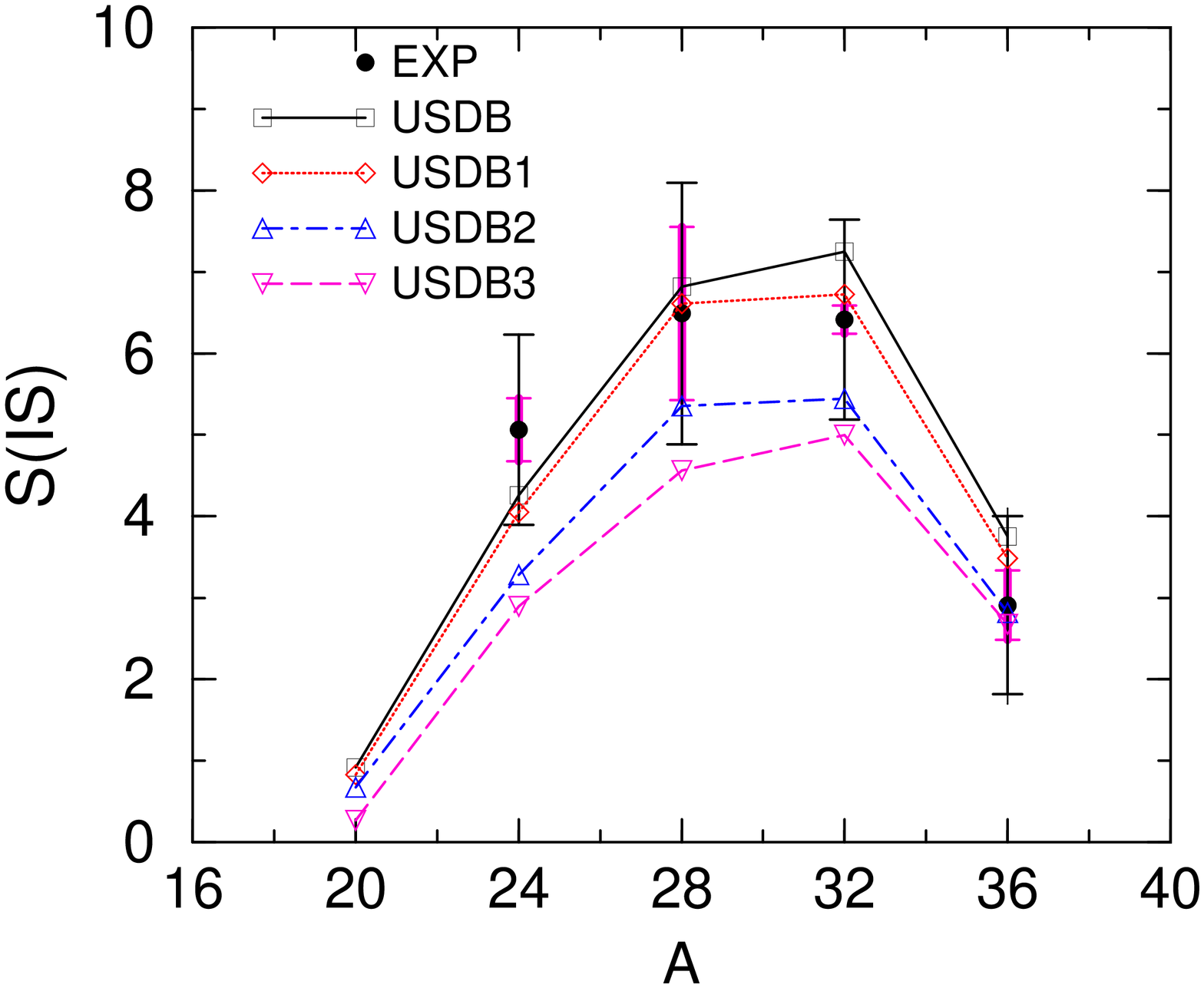}
\includegraphics[scale=0.3,clip,angle=-90
,bb=0 0 595 842]{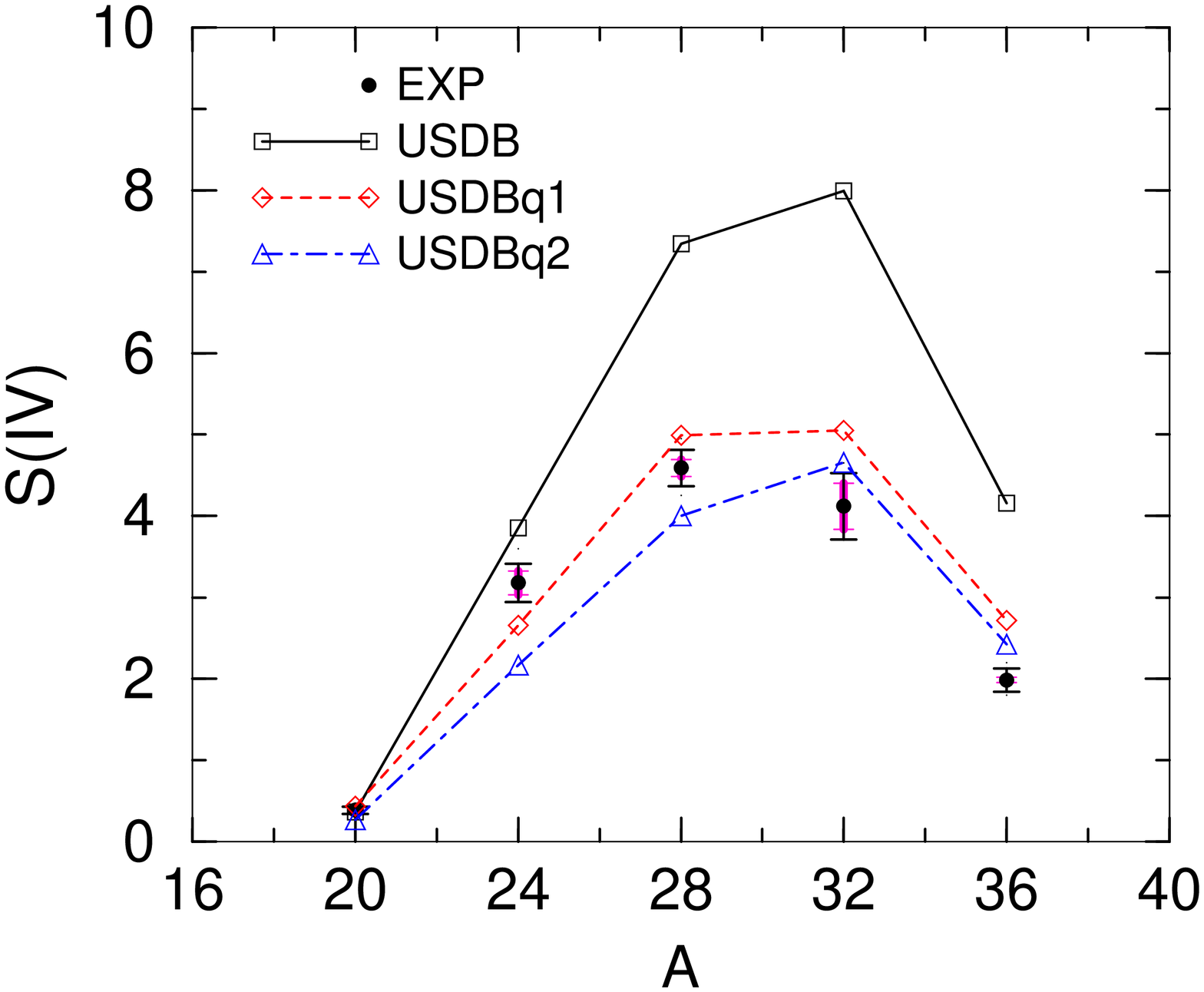}
\vspace{-0.5cm}
\caption{(Color online) Accumulated spin-M1 transition strengths of IS channel (a) and IV channel (b).
Experimental and theoretical data are summed up to E$x=16$MeV .  
Shell model calculations are performed with the USDB effective interaction: 
(a) In the results of USDB1 and USDB2, the IS spin-triplet pairing interaction is enhanced by multiplying the relevant matrix elements by a factor of 1.1 compared to the  original USDB interactions 
with the quenching factor q=1.0 and 0.9 for IS spin operator, respectively.  
For USDB3, the IS pairing interaction is enhanced by a factor of 1.2 with a quenching factor $q$=0.9 for the IS spin operator.
Experimental data are from ref. \cite{Matsu2015}.
Long thin error bars indicate the total experimental uncertainty, while the short thick error bars denote the partial uncertainty from the spin assignment.  
(b) The effective IV  operators are adopted for spin, orbital and spin-tensor operators.
The effective operators are taken from ref. \cite{Towner}.
For the results of USDBq1 and USDBq2, the  IS pairing interaction is enhanced by a factor of 1.1 and 1.2, respectively, with the effective operators.
\label{fig11}}
\end{figure}

\section{Accumulated strength of IS and IV spin M1 excitations}

Figure~\ref{fig11} shows the sum rule values of $S(\vec \sigma)$ and $S(\vec \sigma \tau_z)$  for the variations of  interactions, respectively.  
The 10\%
  enhanced IS pairing in USDB1 give a small quenching effect on the accumulated IS sum rule value about 5\%
, at most 7\%
  in $^{32}$S and  $^{36}$Ar.  With the quenching factor $f_s^{IS}=0.9$ in USDB2, the IS accumulated strength is further decreased by 22-25\%
compared with the original value by USDB interaction.   The  decrease of the IS summed value is going down further by the 20\% 
enhanced IS pairing to be 29-33\% 
in the case of USB3 results.   
Compared with USDB calculations, the  empirical accumulated IS values are 20\%
 enhanced in $^{24}$Mg and gradually quenched from A=28, 0.95, 0.88 and 0.77 in
 $^{28}$Si,  $^{32}$S and  $^{36}$Ar, respectively.  In general, the quenching effect is 
rather small and at most 23\%
 of the original USDB calculations with the bare spin operator.

The IV accumulated sum rule values up to Ex=16MeV are shown in Fig.  ~\ref{fig11}.
The IS pairing interactions are enhances by factors of 1.1 for USDBq1 and 1.2 for USDBq2, respectively, with the effective operators from ref. \cite{Towner}.  The USDBq1 results give 
31-36\% 
quenching sum rule values, while the stronger IS pairing in USDBq2 gives additional quenching of the strength,  i.e.,  41-45\% 
quenching of the summed strength.  The empirical values show also large quenching 
from   33\%
in $^{20}$Ne,  
15\%
in $^{24}$Mg,  27\%
 in $^{28}$Si, 
47\%
in $^{32}$S and 52 \%
in $^{36}$Ar, respectively, compared with the summed value of USDB calculations.  


\begin{figure}[t]
\includegraphics[scale=0.3,clip,angle=-90
,bb=0 0 595 842]{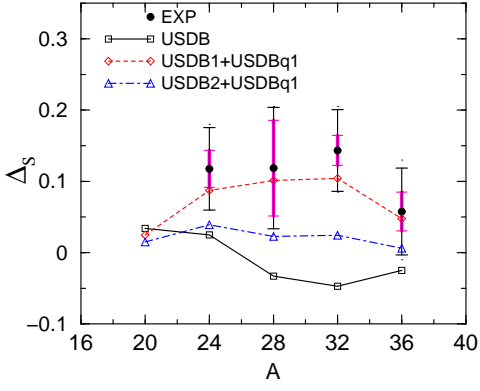}
\caption{(Color online) 
Experimental and calculated proton-neutron spin-spin correlation $\Delta_{spin}$.  
Spin M1 transition strengths are summed up to $E_x$=16MeV.  
Shell model calculations are performed with an effective interaction USDB.  
In the results of USDB1 and USDB2 for the IS channel, the IS spin-triplet interaction is enhanced  multiplying the relevant matrix elements by a factor of 1.1 compared to the original USDB,  and the IS quenching factor is q=1.0 and 0.9, respectively.  The effective spin operators are used for the USDBq1 for the IV transitions. 
Experimental data are taken from ref. \cite{Matsu2015}. 
See the caption of Fig. \ref{fig11} for a description of the experimental error bars. 
\label{fig:spin-spin}}
\end{figure}

Figure \ref{fig:spin-spin} shows the experimental and the calculated proton-neutron spin-spin correlations (\ref{spin-spin}).
Although the experimental data still have large error bars, the calculated results with the USDB interaction show poor agreement with the experimental data.  
This is also the case for the other USD interactions such as USD and USDA.  
The results with an enhanced IS spin-triplet pairing improve the agreement appreciably.
For the IS channel, USDB1 adopts the bare spin operator, while the effective spin operator 
$f_s^{IS}=0.9$ is used for USDB2 case.  The effective operator gives a smaller 
spin-spin correlation $\Delta_{spin}$ than the case of bare IS spin operator.  
The positive value of the correlation indicates that the population of spin triplet pairs in the ground state is larger than that of the spin singlet pairs. We should remind that the spin and the spin-isospin M1 strength may exist in the energy region above Ex=16MeV.  In the analysis
of Figure \ref{fig:spin-spin}, these higher energy contributions are assumed to be the same for both the IS and IV channels.  In the shell model calculations in the full $sd-$shell, the spin M1 strength above 
16MeV is small in  the N=Z nuclei except $^{20}$Ne and $^{28}$Mg.  In $^{20}$Ne,  20\% and 51\% 
of the total strength exist in the energy region Ex=(16-46)MeV for the IS and IV channels, respectively.  In the case of  $^{28}$Mg,  the strengths above Ex=16MeV are  9.9\% and 21\% of the total strengths for IS and IV channels, respectively.    
In the other nuclei, the spin strengths above Ex=16MeV are rather small around few \%
 of the accumulated strength below Ex=16MeV and the strengths are more or less the same in both IS and IV channels. 
The  hypothesis of equal IS and IV strengths  in the higher energy region is valid in the present theoretical calculations in nuclei $A>28$,  which should be checked experimentally in future.

To clarify the physical mechanism of the IS spin-triplet interaction, we make a perturbative treatment of the 2-particle 2-hole (2p-2h) ground state correlations on the spin-spin matrix element.   
We express the wave function for the ground state with proton-neutron correlations for even-even N=Z nuclei, $|\tilde{0}\rangle$, as 
\begin{eqnarray}
&&|\tilde{0}\rangle=|0\rangle 
+\sum_{i=1,1',2,2'}\alpha(1,1',2,2')  \nonumber \\
&\times& |(1_{\pi}1^{'-1}_{\pi})J_1,T_1;(2_{\nu}2_{\nu}^{'-1})J_1,T_1:J=T=0\rangle.
\label{eq:pt1}
\end{eqnarray}
Here the first term on the right hand side, $|0\rangle$, is the wave function with seniority $\nu=0$ (i.e., without the spin-triplet correlations).
The second term represents the states of 2 particle$-$hole pairs $(1_{\pi}1_{\pi}^{'-1})$ and  $(2_{\nu}2_{\nu}^{'-1})$ for proton  and neutron excitations.
The indices $(i\equiv 1,2,1',2')$ stand for the quantum numbers of the single-particle state $i=(n_i,l_i,j_i)$.  
In Eq. (\ref{eq:pt1}), the perturbative coefficient is given by 
\begin{eqnarray}
&&\alpha(1,1',2,2')   \nonumber \\
&=&\frac{\langle (1_{\pi}1^{'-1}_{\pi})J_1,T_1;(2_{\nu}2_{\nu} ^{'-1})J_1,T_1:J=T=0|
H_{p}|0\rangle}{\Delta E}   \nonumber \\
\label{eq:alpha}
\end{eqnarray}
where $H_p$ is the IS spin-triplet two-body pairing interaction and $\Delta E =E_0-E(11^{'-1};22^{'-1})$.
The 2p-2h states are the seniority $\nu$=4 states in Eq. (\ref{eq:pt1}).  
The two-body matrix element in Eq. (\ref{eq:alpha}) is rewritten as
\begin{eqnarray*}
&&\langle (1_{\pi}1^{'-1}_{\pi})J_1,T_1;(2_{\nu}2_{\nu} ^{'-1})J_1,T_1:J=T=0|H_{p}|0\rangle  \nonumber \\
&=&
   \left\{
    \begin{array}{rrr}
      j_{\pi} &   j_{\pi}'  & J_1 \\
      j_{\nu}'& j_{\nu} & J_1' \\
    \end{array}
  \right\}
  \left\{
    \begin{array}{rrr}
      1/2 &  	1/2 & T_1 \\
      1/2 & 1/2 & T_1' \\
    \end{array}
  \right\}          \nonumber \\
&\times&\hat{J}_1\hat{J}_1'\hat{T}_1\hat{T}_1'  (-)^{j_{\pi}'+j_{\nu}+J+J'+1+T+T'} \nonumber \\
&\times&\langle (1_{\pi}2_{\nu})J_1',T_1';(1^{'-1}_{\pi}2_{\nu} ^{'-1})J_1',T_1':J=T=0|H_{p}|0\rangle
\end{eqnarray*}
where $\hat{J}\equiv \sqrt{2J+1}$ and $6j$ symbols are used.  
The matrix element is further expressed as
\bea
&&\langle (1_{\pi}2_{\nu})J_1',T_1';(1^{'-1}_{\pi}2_{\nu} ^{'-1})J_1',T_1':J=T=0|H_{p}|0\rangle   \nonumber \\
&&=-\sqrt{2J_1+1}\langle (1_{\pi}2_{\nu})J_1',T_1'|H_{p}|(\bar{1}^{'}_{\pi}\bar{2}_{\nu} ^{'})J_1',T_1'\rangle
\eea
Since the pairing interaction $H_p$ is attractive and the energy denominator $\Delta E$ is negative,  the perturbative coefficient $\alpha(1,2,1',2')$ should be positive.  
The effect of the ground state  correlations on the proton-neutron spin-spin matrix is then  evaluated as
\begin{eqnarray}
&&\langle \tilde{0}|{\vec S_p}\cdot {\vec S_n} | \tilde{0}\rangle 
= 2\sum_{1,1',2,2'}\alpha(1,2,1',2') \nonumber \\ 
&\times&\langle0|{\vec S_p}\cdot {\vec S_n}|(1_{\pi}1^{'-1}_{\pi})J_1,T_1;(2_{\nu}2_{\nu} ^{'-1})J_2,T_2:J=T=0\rangle  \nonumber \\
\label{eq:pt2}
\end{eqnarray}
where the angular momenta and the isospins are selected to be $J_1=J_2=1$ and $T_1=T_2=0$  by the nature of the neutron-proton spin-spin matrix.
The matrix element in Eq. (\ref{eq:pt2}) is further expressed as a reduced matrix element in the spin space,
\begin{eqnarray}
&&\langle 0||{\vec S_p}\cdot {\vec S_n}||(1_{\pi}1^{'-1}_{\pi})J_1;(2_{\pi}2_{\pi} ^{'-1})J_2:J=0\rangle   \nonumber \\
&=&\delta_{J_1,J_2} \delta_{J_1,1}
\frac{1}{\sqrt{3} } 
(-)^{j_1+j_2-j_1'-j_2'-1}\langle j_1'||{\vec s_p}||j_1\rangle 
 \langle j_2'||{\vec s_n}||j_2\rangle    .  \nonumber \\
\label{eq:redmat}
\end{eqnarray}
In  Eq.(\ref{eq:redmat}), the coupled angular momentum $J'$ is taken as $J'=1$ due to the selection rule of the spin matrix element.  
The isospin quantum number is discarded since it gives a trivial constant in Eq. (\ref{eq:redmat}).
We can obtain the effect of the IS spin-triplet pairing correlations  on  the proton-neutron spin-spin correlation matrix element 
taking a 2p$-$2h configuration $j_1=j_2=j_<=1d_{3/2}$ and  $j_1'=j_2'=j_>=1d_{5/2}$.   Taking the spin-orbit splitting between $1d_{5/2}$ and $1d_{3/2}$ as
5MeV and the spin-triplet pairing matrix element as -2MeV, the $\alpha$ coefficient in Eq. (\ref{eq:alpha}) is evaluated to be 0.056.  The effect on the neutron-proton spin-spin correlation becomes a large positive value, $\Delta_{spin}$=0.27.  
Thus, the positive value obtained by numerical results shown in Fig. 12 can be qualitatively understood by using these formulae for 2p-2h configuration mixing due to the IS spin-triplet pairing.

\section{beta-decay and IS magnetic moments}
The effects of IS pairing and the effective spin operators are  examined for the beta-decay rates between mirror nuclei  with $T=1/2$ and $T_z=\pm1/2$.  The results are given in Table \ref{tab:beta} for USDB, USDB* \cite{Sagawa2016}, USDBq1 and USDBq2 interactions. The IS spin-triplet pairing matrix is enhanced by a factor 1.2 in the relevant matrix elements and bare spin operator is used for USDB*.
\begin{table}
\caption{Beta-decay rate between mirror nuclei  with $T=1/2$ and $T_z=\pm1/2$.   The shell model calculations are performed by using 
USDB and USDB* \cite{Sagawa2016} interactions with the bare spin-g factor as well as by USDBq1 and USDBq2 interactions with the Towner's effective GT operator.}    \label{tab:beta}
\begin{center}
\begin{tabular}{ccccccc} \hline
A  &  J  &  USDB & USDB* & USDBq1 & USDBq2 & Exp.   \\  \hline 
19  &  1/2 &    2.761 &   2.800 &  2.076 & 2,096 & 1.652  \\
21  &  3/2  &   0.486 &  0.514  & 0.372 & 0.398 & 0.323  \\
23 &  3/2   & 0.267&   0.230  & 0.200 & 0.183 & 0.190   \\
25  &  5/2  &  0.616 &  0.661  & 0.515 & 0.539 & 0.414   \\
27  & 5/2  &  0.464  & 0.427  & 0.404 & 0.363 & 0.304   \\
29  & 1/2  &  0.233  & 0.180  & 0.194 & 0.154 & 0.176    \\
31  & 1/2  &  0.288 &  0.242  & 0.216 & 0.196 & 0.176    \\
33  & 3/2  &  0.171 &  0.156  & 0.135 & 0.127 & 0.0577    \\
35  & 3/2  &  0.138 &  0.143  & 0.116 & 0.117 & 0.0505    \\
37  & 3/2 &   0.386 &  0.419 & 0.315 & 0.332 & 0.218     \\\hline 
  & $\sigma_{rms}$  &0.37 & 0.39& 0.15 & 0.16 &    \\\hline 
\end{tabular}  
\end{center}
\end{table}
 The r.m.s. deviations $\sigma_{rms}$ of the calculations and experiments are 0.37, 0.39, 0.15 and 0.16 for USB, USDB*, USDBq1 and USDBq2, respectively.
We do not see any significant difference in the results between USDB and USDB* interactions.  

 Let us make the optimal $\chi-$ square fits with experimental data  by using the quenching  spin $q$ factor  for the bare GT operator.
The  r.m.s. deviation  $\sigma_{rms}$  of $q_{GT}(eff)^2\times$B(GT; USDB) from B(GT; EXP) becomes the minimum value 0.030 at $q_{GT}(eff)$=0.78, while
that of $q_{GT}(eff)^2\times$B(GT; USDB*) from B(GT; EXP) becomes the minimum value 0.039 at $q_{GT}(eff)$=0.77.  
We obtain more or less the same quenching factor for USDB and USDB*.  
This is because odd nucleon in these T=1/2 nuclei masks the pairing effect by the single-particle nature of the transition.
The r.m.s deviations $\sigma_{rms}$  get reduced for the cases with the Towner's effective operator, USDBq1 and USDBq2, though they are not as small as those for the $\chi-$square fitted cases with the quenching factor $q_{GT}(eff)$.  

 The traditional source of information for the IS spin operator is from isoscalar magnetic moments, i.e., the average of the magnetic 
moments of mirror nuclei.  The IS magnetic moments are also calculated for USDB, USDB* and USDB3 as well as for the case with Towner's effective operator (Eq. (9)) and tabulated in Table \ref{tab:ISmm}.
In case for the Towner's operator which will be denoted as 'Towner', the IS spin-triplet pairing matrix is not enhanced. 
The case for using the effective operator (Eq. (9)) where the coefficients are obtained by $\chi-$square fitting the experimental isoscalar spin expectation value of $sd-$shell nuclei with $T$ =1/2, 0 and 1 (TABLE II of Ref. \cite{BW1983}). The coefficients  obtained are $f_s^{IS}$ =0.66, $f_l^{IS}$ =0.05 and $f_p^{IS}$ =0.06 and the r.m.s. deviation  $\sigma_{rms}$  is 0.044.  The results using enhanced IS spin-triplet pairing matrix by a factor 1.2 is also tabulated  denoted as USDB*$_{eff}$. 
The $\sigma_{rms}$  of the calculations and experiments shown in Table \ref{tab:ISmm} are 0.032, 0.033, 0.030, 0.020 and 0.019 for USB, USDB*, USDB3, Towner and USDB*$_{eff}$, respectively.
The results of Towner and USDB*$_{eff}$ are close to each other with smaller r.m.s deviations and quite satisfactory.
We do not see any appreciable difference among the results of USDB, USDB* and USDB3, whose r.m.s deviations are larger but only 1.5 times those of Towner and USDB*$_{eff}$. 

The IS magnetic moment is expressed by using the effective operators as \cite{BW1983}
\be
[\mu_{IS}-J/2]/(g_s^{IS}-g_l^{IS})=\langle J|s_z|J\rangle+g_s^{IS}/(g_s^{IS}-g_l^{IS})\delta s
\label{IS-mm}
\ee
where
\be
\delta s=\delta_s\langle J|s_z|J\rangle+\delta_l\langle J|l_z|J\rangle+\delta_p\sqrt{8\pi}\langle J|[Y_2\times\vec{s}]^{(1)}_0|J\rangle.
\label{eq:spin-q}
\ee
The effective factors $\delta_s$,  $\delta_l$ and $\delta_p$ are essentially equivalent to 
the effective factors  in Eq.  (\ref{eq:IS-eff}) as $\delta_s=f_s^{IS}-1$, $\delta_l=f_l^{IS}/2$ and $\delta_p=f_p^{IS}$ .  
Since the effects of spin and orbital contributions in Eq.  (\ref{eq:spin-q}) to the IS magnetic moment cancel
largely because of different signs of $\delta_s$ and  $\delta_l$ so that the net effect of the effective operators  is rather small.

 \begin{table}
\caption{IS magnetic moments extracted from the average of magnetic moments in mirror nuclei.  The shell model calculations are 
performed by using USDB, USDB* and USDB3 interactions as well as for the case with the Towner's effective operator (denoted as Towner). 
The case with the effective operator (Eq. (9)) where the coefficients are obtained by chi-square fitting procedure (denoted as USDB*$_{eff}$) is also given.
The values are given in a unit of nuclear magneton 
$\mu_N$}.  \label{tab:ISmm}

\begin{center}
\begin{tabular}{cccccccc} \hline
A  &  J  &  USDB & USDB*  & USDB3 & Towner & USDB*$_{eff}$ & Exp.   \\  \hline 
19 & 1/2  & 0.431  &0.432 & 0.414 & 0.385 & 0.371 & 0.371  \\
21  &3/2  & 0.869  &0.870 & 0.858 & 0.850 & 0.853 &0.862\\
23  &3/2  & 0.844  &0.838 & 0.829 & 0.832 & 0.834 & ---\\
25  &5/2  & 1.403 & 1.403 & 1.388 & 1.384 & 1.397 & 1.395\\
27  &5/2  & 1.388 & 1.381 & 1.368 & 1.373 & 1.384 & 1.393\\
29 & 1/2  & 0.315 & 0.304 & 0.298 & 0.301 & 0.297 & 0.340\\
31  &1/2  & 0.323 & 0.313 & 0.307 & 0.307 & 0.302 & 0.322\\
33  &3/2  & 0.691 & 0.676 & 0.683 & 0.711 & 0.722 & ---\\
35 & 3/2   &0.676&  0.679 & 0.686 & 0.715 & 0.724 & 0.727\\
37 & 3/2  & 0.629  &0.630 & 0.642 & 0.681 & 0.692 & ---\\\hline 
& $\sigma_{rms}$ & 0.032 &0.033&0.030&0.020&0.019 & \\\hline
\end{tabular}  
\end{center}
\end{table}
 \begin{table*}
\caption{Summed IS and IV spin M1 transitions in $s-d$ shell nuclei and $^{12}$C. Calculated and experimental values are accumulated up to Ex=16MeV. 
The shell model calculations are 
performed by using USDB and USDB*(the IS pairing is enhanced)  
 interactions
 as well as for the case with the Towner's effective operator (denoted as Towner) for USDB results.  For $^{12}$C, the shell model calculations are performed with  SFO interaction with the bare spin operator.  The experimental IS strength is for the state at Ex=12.708MeV, while the IV strength is for the state at Ex=15.113MeV.  Experimental data are taken from ref. \cite{Matsu2015}. The values in the bracket are experimental errors.  
\label{tab:IS-IV-S}
}
\begin{center}
\begin{tabular}{cccccccccc} \hline
 &\multicolumn{4}{c}{S(IS)} &\multicolumn{4}{c}{S(IV)}  \\ \hline
A  &   USDB & USDB*  & Towner &  Exp. & USDB & USDB* &Towner & Exp.   \\  \hline 
$^{20}$Ne  & 0.919 & 0.574  &0.485 &--- & 0.577& 0.346 & 0.456 & 0.387 (0.042)  \\
$^{24}$Mg   &4.256 & 3.573  &2.221 & 5.05 (1.00) & 3.856& 2.937 & 2.808 & 3.3 (0.3) \\
$^{28}$Si   &6.816 &  5.632 & 3.579 &  6.5 (1.5) & 7.340  & 5.593  & 5.256  & 4.6 (0.35)\\
$^{32}$S  & 7.247  & 6.167 & 3.810 & 6.4 (1.2)  & 7.993  & 6.554 & 5.734  & 4.2 (0.45) \\
$^{36}$Ar   & 3.753 & 3.303& 1.975& 2.9 (1.1) & 4.159  & 3.414 & 2.978 & 2.0 (0.2)\\ \hline 
$^{12}$C  &  1.516  & ---  &---  & 3.174(0.842) &  1.937 & --- & --- & 1.909 (0.094)\\
\hline 
\end{tabular}  
\end{center}
\end{table*}
 The summed IS and IV spin M1 transitions are tabulated in Table \ref{tab:IS-IV-S} for N=Z $s-d$ shell nuclei and $^{12}$C.  
The calculated values are obtained by the USDB and USDB* interactions with the bare spin operator as well as with the Towner's effective operators both for IS and IV transitions.
The values of  $^{12}$C are for the IS state at Ex=12.708MeV and for the IV one at Ex=15.113MeV, respectively.  For the summed strength,  the enhanced IS interaction USDB* 
gives 12-18\% 
quenching  for the IS spin transitions   and 18-24\% 
 quenching for the IV ones  except $^{20}$Ne (38\%
for IS and 40\%
for IV).   The different values between $^{20}$Ne and the other nuclei is due to the smaller summed values below Ex=16MeV for $^{20}$Ne,  which exhaust only 80\%
 and half of the total strength of full model space for IS and IV channels, respectively, while  more than  90\%
of the total strength exists  below Ex=16MeV  in other nuclei (except IV channel of $^{28}$Mg; 80\% ).    The Towner's effective operators give large quenched values, about 47-48\% 
 quenching for  all nuclei in the IS channel and 27-28\% 
quenching (21\% for $^{20}$Ne) 
 for the IV one.   
The experimental summed values of the IS strengths show rather minor quenching or even enhanced in $^{24}$Mg  compared with USDB results, while the IV data show a large 
quenching consistent with those obtained by Towner's effective operators.  
More quantitatively,  we need a combined effect of the enhanced IS paring and the effective operators to get better agreement with the experimental values of IV channel for A$\geqq$28 as seen in Fig. 11. 
Although the IS moments are very well described by Towner and USDB*$_{eff}$, the calculated $S(IS)$ values prove to be much suppressed compared with the experimental data. The values of accumulated strengths  $S(IS)$ are 0.485 (0.318), 2.221 (1.520), 3.579 (2.356), 3.810 (2.490) and 1.975 (1.282) for $^{20}$Ne, $^{24}$Mg, $^{28}$Si, $^{32}$S and $^{36}$Ar, respectively, for Towner's operators (USDB*$_{eff}$).

 The difference between the IS magnetic moment and the spin M1 transition will be 
clarified by the following way.  The IS spin M1 transition is expressed as
\be
\langle J|\hat{O}(\sigma)|0\rangle=(f_s^{IS}-f_l^{IS})\langle J|\vec{\sigma}|0\rangle+f_g^{IS}\sqrt{8\pi}\langle J|[Y_2\times\vec{\sigma}]^{(1)}|0\rangle
\ee
where the non-diagonal matrix element of the operator $\vec{J}$ vanishes,   and the effective operators $f_s$ and $f_l$ have different signs so that the net effect gives a large quenching effect.  This is rather different from  the IS magnetic moment in Eq. (\ref{IS-mm}) where the effects of 
effective  spin and orbital operators cancel largely. 

\section{Summary} 

In summary, we studied the IS and IV spin M1 transitions in even-even N=Z $sd-$shell nuclei using shell model calculations with  USDB interactions in full $sd-$shell model space. 
We introduced the effective operators for the spin and spin-isospin M1 operators in Eqs. (9) and (11) as well as the enhanced IS spin-triplet pairing. 
In general, the calculated results show good agreement with the experimental energy spectra in N=Z nuclei as far as the excitation energies are concerned. 
 Compared with the experimental M1 results,  the accumulated IS spin strengths up to 16MeV show small quenching effect, corresponding to the effective quenching 
operator $q^{IS}(eff)\sim$0.9, while  a large quenching $q^{IV}(eff)\sim$0.7 is extracted for the IV channel.  
The similar quenching on the IS spin M1 transitions is  obtained by  the 20\% enhanced  IS spin-triplet pairing correlations 
with the bare spin operator. 
 The enhanced IS pairing does not change much  the excitation energy spectra themselves. 
Positive contributions for the spin-spin correlations are  found by an enhanced isoscalar spin-triplet pairing interaction in these $sd-$shell nuclei.  
The effects of the effective spin operators and enhanced IS pairing is also examined on the beta decay rate and the IS magnetic moments in $sd-$shell nuclei.  The r.m.s. deviation between the calculations and experiments of beta decay rate are improved by the effective operator, while the IS pairing  has only a minor effect on these observables since the unpaired particle masks the effect of pairing correlations on the matrix elements. 

The Towner's effective spin operators works well to reproduce the accumulated experimental IV spin strength, while the quenching of the effective operators is much larger than the observed one in the IS spin channel.  In the past, a large quenching of IS magnetic transition 
strength was suggested in the literatures.  However, the (p,p') data in ref. \cite{Matsu2015} do not show any sign of the large quenching effect on the IS spin transitions.  This point should be studied further experimentally by possible IS probes such as $(d,d')$ reactions \cite{Kawa2004} together with comprehensive theoretical calculations.  
\begin{acknowledgments}
\vspace{-0.3cm}
We would like to thank  H. Matsubara for providing the experimental data.
We would also like to thank M. Ichimura,  A. Tamii, M. Sasano and T. Uesaka for the useful discussions.
This work was supported in part by JSPS KAKENHI  Grant Numbers JP16K05367 and JP15K05090.
\end{acknowledgments}

\end{document}